\documentclass[aps,amsmath,amssymb,reprint,superscriptaddress]{revtex4-2}

\usepackage{graphicx}% Include figure files
\usepackage{dcolumn}% Align table columns on decimal point
\usepackage{bm}% bold math
\usepackage{ulem} % strikethrough
\usepackage{siunitx} % Units, micrometer in particular
\renewcommand{\Im}{\mathord{\mathrm{Im}}}

\begin{document}

\preprint{APS/123-QED}

\title{Dynamical Backaction Evading Magnomechanics}% Force line breaks with \\
%\title{Triple-Resonance Backaction Evasion}

\author{C.A. Potts}
\email{c.a.potts@tudelft.nl}
\affiliation{Department of Physics, University of Alberta, Edmonton, Alberta T6G 2E9, Canada}

\author{Y. Huang}
\affiliation{Department of Physics, University of Alberta, Edmonton, Alberta T6G 2E9, Canada}

\author{V.A.S.V. Bittencourt}%
\affiliation{Max Planck Institute for the Science of Light, Staudtstr. 2, 91058 Erlangen, Germany}%

\author{Silvia {Viola Kusminskiy}}
\affiliation{Institute for Theoretical Solid State Physics, RWTH Aachen University, 52074 Aachen, Germany}
\affiliation{Max Planck Institute for the Science of Light, Staudtstrasse 2, 91058 Erlangen, Germany}

\author{J.P. Davis}
\email{jdavis@ualberta.ca}
\affiliation{Department of Physics, University of Alberta, Edmonton, Alberta T6G 2E9, Canada}

\date{\today}% It is always \today, today,

\begin{abstract}
The interaction between magnons and mechanical vibrations dynamically modify the properties of the mechanical oscillator, such as its frequency and decay rate. Known as dynamical backaction, this effect is the basis for many theoretical protocols, such as entanglement generation or mechanical ground-state cooling. However, dynamical backaction is also detrimental for specific applications. Here, we demonstrate the implementation of a triple-resonance cavity magnomechanical measurement that fully evades dynamical backaction effects. Through careful engineering, the magnomechanical scattering rate into the hybrid magnon-photon modes can be precisely matched, eliminating dynamical backaction damping. Backaction evasion is confirmed via the measurement of a drive-power-independent mechanical linewidth. 
\end{abstract}

\maketitle

%\tableofcontents 

\begin{figure}[b]
\includegraphics{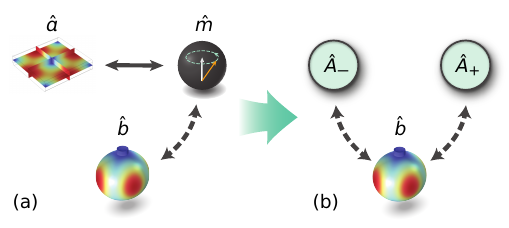}
\caption{Schematic of the hybrid magnomechanical system. (a) Numerical simulation of the bare magnetic field distribution of the microwave resonance, $\hat{a}$, schematic representation of the uniform Kittel magnon mode, $\hat{m}$, numerical simulation of the mechanical breathing mode of the YIG sphere, $\hat{b}$. (b) Due to strong coupling between the magnon and photon modes this system can be recast, see main text, in terms of a single mechanical mode coupled to two independent bosonic normal modes --- the fundamental requirement for the dynamical backaction evasion developed here.}
\label{Fig:01}
\end{figure}

When a mechanically compliant mirror reflects light, momentum is transferred to the mirror generating a force known as radiation pressure \cite{ashkin1980applications,cohadon1999cooling}. Within an optical resonator, each photon takes many round trips before it decays due to the finite cavity lifetime, enhancing the radiation pressure force \cite{metzger2004cavity}. This enhancement has enabled the dynamical modification of the properties of a coupled mechanical element \cite{aspelmeyer2014cavity,chan2011laser,teufel2011sideband}, known as dynamical backaction. Similarly, magnetic excitations in solids (magnons) \cite{chumak2021roadmap,rameshti2022cavity,lachance2019hybrid} have been demonstrated to impart a radiation pressure-like force on the mechanical vibrations of the magnetized material \cite{zhang2016cavity,potts2021dynamical,shen2022mechanical}. Typically, in such experiments, a magnetic sphere of yttrium iron garnet (YIG) \cite{FerriSphere} is loaded into a microwave cavity, which allows both drive and measurement of magnons \cite{huebl2013high,zhang2014strongly,tabuchi2015coherent,bourcin2022strong,potts2020strong,wang2019nonreciprocity,harder2018level,zhang2015magnon,morris2017strong,streib2019magnon}. The uniform magnetic excitation, called the Kittel mode \cite{walker1958resonant,fletcher1959ferrimagnetic}, couples simultaneously to the microwave cavity mode and to the mechanical vibrations \cite{Schlomann1960generation,keshtgar2014acoustic,callen1968magnetostriction}, see Fig.~\ref{Fig:01}(a).

When the magnons are driven, dynamical backaction manifests as a modification in the frequency and decay rates of the vibrations, referred to as the magnon spring effect and magnomechanical damping, respectively. These phenomena were predicted \cite{potts2020magnon} and subsequently demonstrated in a cavity magnomechanical system \cite{potts2021dynamical}. The experimental realizations of cavity magnomechanical systems and the above-mentioned backaction effects have stimulated many theoretical proposals, such as the generation of entanglement \cite{fan2022microwave,fan2022microwave,li2021entangling,li2019entangling,li2018magnon}, squeezed states \cite{li2021squeezing,ding2022magnon,asjad2022magnon,li2019squeezed,zhang2021generation}, and quantum and classical information processing \cite{sarma2021cavity,li2020phase,kong2019magnetically,li2021quantum,hatanaka2022chip}. However, while dynamical backaction is a powerful tool and an essential part of such proposals, it is also often undesired.

Notably, dynamical backaction will displace a mechanical vibrational mode from thermal equilibrium with its environment \cite{clark2017sideband}. If one wishes to ensure the mechanical mode is in thermal equilibrium with its environment, dynamical backaction must be avoided. This is critical, for example, to use cavity magnomechanics as a primary thermometer \cite{purdy2017quantum,potts2020magnon}, or to observe quantum backaction effects \cite{hertzberg2010back,teufel2016overwhelming,cripe2019measurement,purdy2013observation}.

\begin{figure}[t]
\includegraphics[width = 0.45\textwidth]{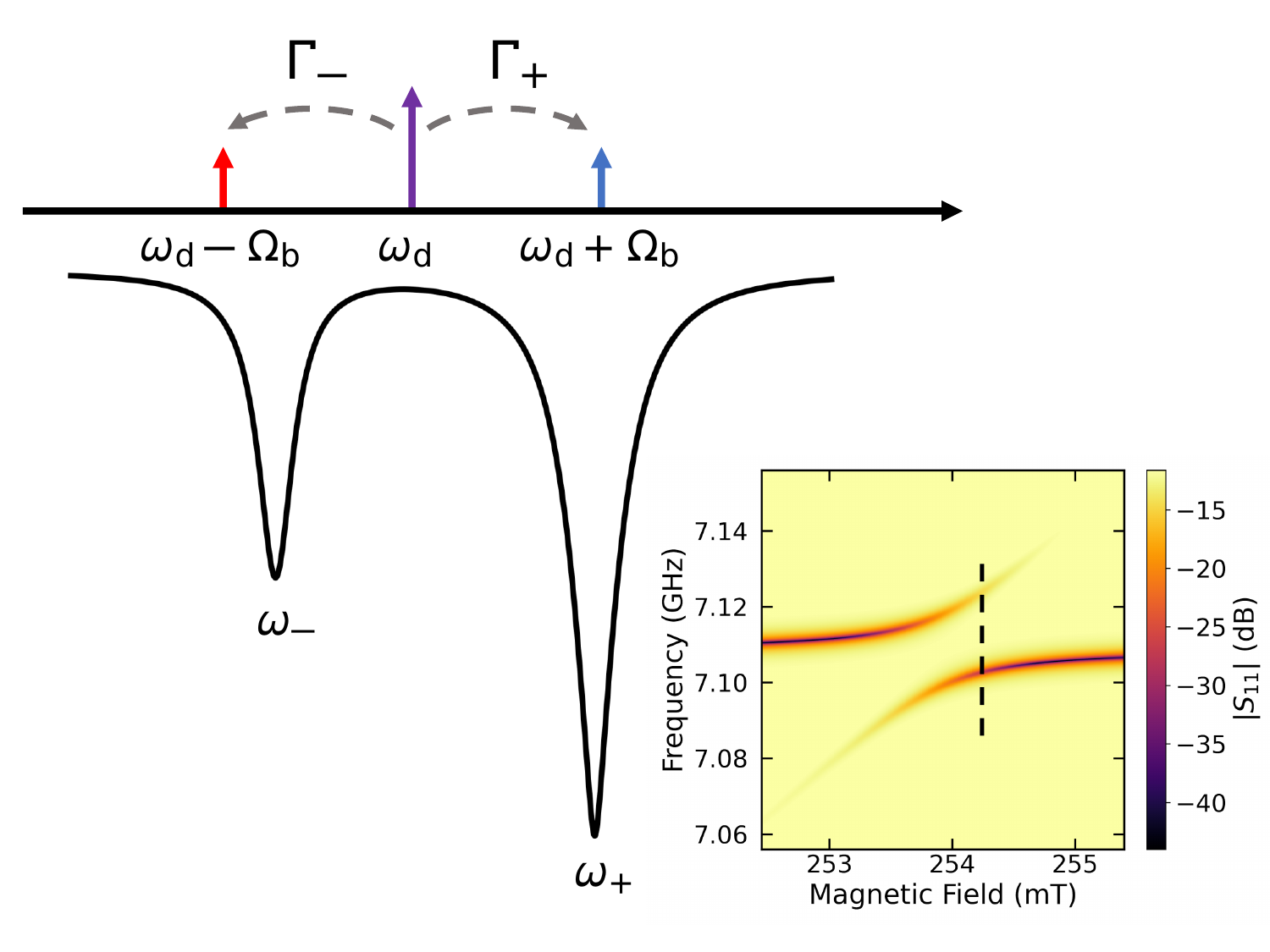}
\caption{Schematic of the scattering picture of a cavity magnomechanical system.} A coherent microwave drive tone, tuned between the normal modes, results in simultaneous Stokes and anti-Stokes scattering processes. Balancing the rates of these processes, $\Gamma_-$ and $\Gamma_+$, results in the evasion of dynamical backaction. Inset: Measured normal mode spectrum as a function of the applied magnetic field. The dashed line corresponds to the spectrum shown in the figure.
\label{Fig:02}
\end{figure}

This article reports on the experimental realization of a cavity magnomechanical measurement free of magnomechanical damping, referred to as dynamical backaction evasion. We use a two-phonon triple resonance condition, where the frequency difference between the hybrid cavity-magnon modes is approximately twice the phonon frequency, Fig.~\ref{Fig:02}. By carefully setting the drive frequency, the magnomechanical scattering rate into the two hybrid modes formed due to the strong magnon-microwave interaction match, cancelling dynamical backaction effects. We demonstrate the ability to smoothly tune our system from magnomechanical cooling (enhancement of the decay rate), through complete backaction evasion, to a regime of magnomechanical amplification (reduction of the decay rate). In this process, we observe a shift in the expected magnomechanical decay rate from that predicted by the linear theory, which we attribute to the intrinsic Kerr nonlinearity of YIG and coupling of the phonon mode to weakly driven Walker modes \cite{gloppe2019resonant,BittencourtDamping2022}. This work represents a critical milestone towards the implementation of practical quantum correlation thermometry \cite{potts2020magnon} and is expected to lead to the direct observation of magnon-induced quantum backaction \cite{braginsky1980quantum,braginsky1995quantum}.

The dynamical backaction effects in a cavity magnomechanical system can be described by a linearized theory outlined in Ref.~\cite{potts2020magnon}. The semi-classical steady state of the system under the external microwave drive is evaluated, and the dynamics of the fluctuations are described by a quadratic Hamiltonian. However, it is instructive to recast such a Hamiltonian in terms of the hybrid magnon-microwave normal modes, see Fig.~\ref{Fig:01}(b), which is outlined in the Supplementary Material \cite{SupplimentaryInfo}, and given by
\begin{equation}
\begin{aligned}
    \hat{\mathcal{H}} =&{}  -\hbar\Delta_+ \hat{A}_+^{\dagger}\hat{A}_+ - \hbar\Delta_- \hat{A}_-^{\dagger}\hat{A}_- + \hbar\Omega_{\rm{b}} \hat{b}^{\dagger}\hat{b} \\
    &{}+ \hbar g_+ (\hat{A}_+^{\dagger} + \hat{A}_+)(\hat{b} + \hat{b}^{\dagger})  \\&{}+ \hbar g_- (\hat{A}_-^{\dagger} + \hat{A}_-)(\hat{b} + \hat{b}^{\dagger}).
    \label{eqn:03}
\end{aligned}
\end{equation}
Here, $\hat{b}$ is the annihilation operator for the phonon mode with frequency $\Omega_{\rm{b}}$ and $\hat{A}_+$ and $\hat{A}_-$ are the annihilation operators for the upper and lower normal modes, respectively. Furthermore, $\Delta_{\pm} = \omega_{\rm d} - \omega_{\pm}$ are the drive detuning from the normal modes, see Fig.~\ref{Fig:02}. The parametrically enhanced magnomechanical coupling rates are $g_+ = g_\text{mb}^0 [\langle \hat{A}_+ \rangle\sin^2\theta + \langle \hat{A}_- \rangle\sin (2 \theta) /2]$ and $g_- = g_\text{mb}^0 [\langle \hat{A}_- \rangle\cos^2\theta + \langle \hat{A}_+ \rangle\sin (2 \theta) /2]$ for the upper and lower normal modes, respectively, and $g_\text{mb}^0$ is the single-magnon magnomechanical coupling rate. The mixing angle describing the normal mode composition in terms of the magnon and microwave modes is given by $\theta = \tfrac12 \arctan(-2g_{\rm am}/\Delta_{\rm am})$, which depends on the magnon-photon coupling rate $g_{\rm am}$, and the magnon-photon detuning, $\Delta_{\rm am} = \omega_{\rm a} - \omega_{\rm m}$; where, $\omega_{\rm a,m}$ are the cavity and magnon frequencies, respectively. The cavity frequency is defined by the cavity design, whereas the magnon frequency can be modified by an external bias magnetic field \cite{tabuchi2015coherent}. This allows tunability of the frequency splitting between the normal modes. Additionally, $\langle \hat{A}_{\pm} \rangle$ are the steady-state amplitudes of the upper and lower normal modes. Written in this form, the magnomechanical interaction can be understood as a single mechanical mode simultaneously interacting with two `independent' bosonic modes.

In the resolved sideband limit, i.e.\ $ \Omega_{\rm b} \gg \kappa,\gamma_{\rm m}$ --- where $\kappa$ and $\gamma_{\rm m}$ are the decay rates of the microwave cavity and magnon mode, respectively --- and in the weak coupling limit, i.e.\ $g_{\pm} \ll \kappa,\gamma_{\rm m}$, the magnomechanical scattering rates into the upper and lower normal modes are given by
\begin{equation}
    \Gamma_{\pm} = \frac{4 \vert g_{\pm} \vert^2 \kappa_{\pm}}{4(\Delta_{\pm}\pm\Omega_{\rm b})^2+\kappa^2_{\pm}},
\end{equation}
 where $\kappa_{\pm}$ are the effective decay rates of the hybrid modes.

As a result of the magnomechanical interaction, the linewidth of the mechanical oscillator will be modified from its intrinsic value, $\Gamma_{\rm b}$ \cite{potts2021dynamical}. This modification is a consequence of dynamical backaction, and the total mechanical linewidth will be given by $\Gamma_{\rm tot} = \Gamma_{\rm b} + \Gamma_{\rm mag}$, where
\begin{equation}
\label{Eq:GammaMag}
    \Gamma_{\rm mag} \simeq \Gamma_+ -\Gamma_-.
\end{equation}
Therefore, because of the coupling to the two bosonic modes, it is possible to avoid dynamical backaction effects by tuning the system such that the scattering rates into the upper and lower normal modes exactly match, i.e.\ $\Gamma_+ = \Gamma_-$. We should note that such a condition will \textit{not} correspond to the evasion of quantum backaction effects \cite{braginsky1980quantum,hertzberg2010back,teufel2016overwhelming,lecocq2015quantum,clerk2008back,suh2014mechanically,shomroni2019optical,ockeloen2016quantum}). Quantum backaction evasion is often accomplished using a stroboscopic measurement technique described in Ref.~\cite{clerk2008back}, whereas we perform a continuous measurement of the mechanical oscillator.  The most straightforward way for evading dynamical backaction in the zero magnon-photon detuning limit ($\Delta_{\rm am} = 0$) is by engineering the normal mode splitting to be twice the phonon frequency, implying that $g_{\rm am} = \Omega_{\rm b}$ --- the two-phonon triple resonance condition --- and applying a coherent drive tone tuned at $\Delta_- = -\Delta_+ = \Omega_{\rm b}$. In practice, this can be challenging; however, it is generally possible to find a detuning condition such that dynamical backaction evasion is possible, as shown in the experiment described below. We present a discussion on why backaction evasion is not possible in a triply-resonant system such as Ref.~\cite{potts2021dynamical} in the Supplementary Material \cite{SupplimentaryInfo}.

\begin{figure}[t]
\includegraphics{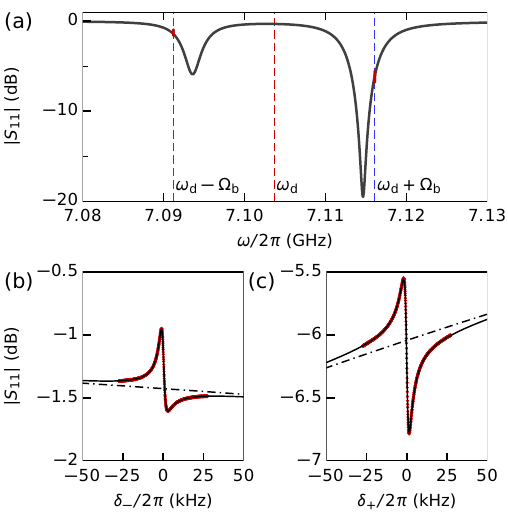}
\caption{Magnomechanically induced transparency/absorption. (a) Normal mode spectrum detuned for dynamical backaction evasion, $\Delta\omega/2\pi = 21.0$ MHz and drive detuning $\Delta_+/2\pi = -12.02$ MHz. Simultaneously induced magnomechanical transparency and absorption peaks (red) are visible in the upper and lower normal modes, respectively. The red dashed line at $\omega_\textrm{d}$ indicates the location of the coherent drive tone. Blue dashed lines indicate the location of the resulting mechanical sidebands. (b) Zoom in on the MMIA peak of the lower normal mode of the panel (a). (c) Zoom in on the MMIT peak of the upper normal mode of the panel (a). Dotted red curves are experimental data; solid grey curves are numerical fits, for brevity, we have defined $\delta_{\pm} = \omega_{\rm d} \pm \Omega_{\rm b}$.}
\label{Fig:03}
\end{figure}

In our experimental setup $g_{\rm am}/2\pi = 9.34$ MHz and $\Omega_{\rm b}/2\pi = 12.45$ MHz. Notice that the system parameters do not fulfill the two-phonon triple resonance condition at maximal hybridization; therefore, dynamical backaction evasion can not be accomplished in our setup by driving with a detuning equal to the phonon frequency. Instead, to compensate for the mismatch between the magnon-photon coupling and the phonon frequency, the magnon mode frequency can be tuned out of resonance from the microwave cavity, $\Delta_{\rm am} \neq 0$. 

The flexibility of the hybrid magnomechanical system allows us to cope with such a mismatch by tuning the normal mode splitting. Here, the normal mode splitting was tuned to a value such that a drive at a specific frequency between the normal modes induces dynamical backaction evasion. Such drive frequency can be estimated by setting Eq.~\eqref{Eq:GammaMag} to zero and solving for the drive frequency. We chose to use a normal mode splitting of $\Delta\omega = \omega_+ - \omega_- =  2\pi \times 21.0$ MHz, see Fig.~\ref{Fig:03}(a); however, this choice was largely arbitrary since all detunings below a specific value enable backaction evasion. For {our} experimental realization, the normal mode splitting cutoff was at approximately $\Delta\omega = 2\pi \times 21.5$ MHz.

\begin{figure}[t]
\includegraphics{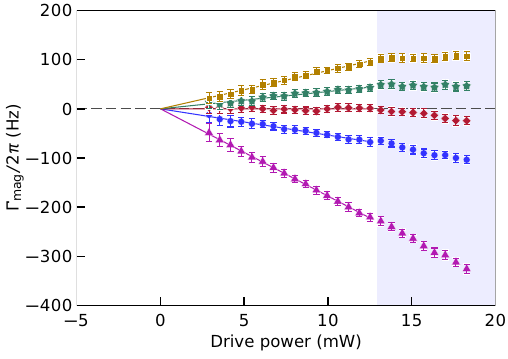}
\caption{Magnomechanical damping rate. Magnomechanical damping rate as a function of power for several drive detunings. For a drive detuning of $\Delta_+/2\pi = -12.02$ MHz (in red), there exists no modification of the total mechanical linewidth and, therefore, no dynamical backaction. The shaded region deviates from the linear behavior because of the Kerr nonlinearity of YIG \cite{wang2018bistability,shen2022mechanical}. The drive detunings here are: $\Delta_+ / 2\pi = $ $-13.40$ MHz (yellow), $-12.40$ MHz (green), $-12.02$ MHz (red), $-11.67$ MHz (blue) and $-11.00$ MHz (magnet).}
\label{Fig:04}
\end{figure}

To probe the dynamical backaction, the mechanical spectrum was measured using magnomechanically induced transparency/absorption (MMIT/MMIA). MMIT was first observed by Zhang $\textit{et~al.}$ \cite{zhang2016cavity} and is the result of interference between the weak probe and the up-converted excitations via the annihilation of a phonon. Applying the coherent drive tone detuned from the upper normal mode, we simultaneously observe MMIT in the upper normal mode and MMIA in the lower normal mode, see Fig.~\ref{Fig:03}. We should note that simultaneous transparency and absorption measurements have previously been observed in optomechanics using two independent drive tones \cite{lake2020two}, whereas our protocol uses a single drive tone.

Performing a fit to the transparency/absorption windows, we can extract the total linewidth of the mechanical oscillator, given by the sum of the intrinsic linewidth and the magnomechanical damping, $\Gamma_{\rm tot} = \Gamma_{\rm b} + \Gamma_{\rm mag}$. The intrinsic linewidth can be obtained by varying the power of the coherent drive and extrapolating to zero power, as shown in Fig.~\ref{Fig:04}. The intrinsic mechanical linewidth is $\Gamma_{\rm b}/2\pi = 3745 \pm 6$ Hz. Varying the drive detuning, $\Delta_+$, it is possible to transition smoothly from magnomechanical damping to magnomechanical anti-damping. For a drive detuning of $\Delta_+/2\pi = -12.02$ MHz, we observe no change in the mechanical linewidth with changing drive power (see Fig.~\ref{Fig:04}), indicating the complete evasion of dynamical backaction.

We observe a departure from the linear behavior at high drive powers ($\mathcal{P} \gtrsim 13$ mW at the cavity) \cite{LeCraw1958nonlinear}. This can be attributed to the Kerr bistability modifying the phonon frequency, as described by Shen \textit{et~al.} in Ref.~\cite{shen2022mechanical}. In addition, at higher drive powers, one would expect a departure due to unavoidable quantum backaction \cite{braginsky1980quantum}. The shaded region in Fig.~\ref{Fig:04} indicates the deviation from linear behavior. The nonlinear behavior is more noticeable for large drive detunings; at these detunings, the coherent drive generates a large steady-state magnon population resulting in a considerable shift of the phonon frequency~\cite{shen2022mechanical}. 

Finally, we compare our results with the linear theory derived in Ref.~\cite{potts2020magnon}. The magnomechanical damping rate is given by $\Gamma_{\rm mag} = 2 \Im \Sigma[\omega]$, where $\Sigma[\omega]$ is the mechanical self-energy that encodes the modification of the mechanical susceptibility due to the magnon-phonon coupling. It is obtained by solving the linearized equations of motion describing the coupled dynamics of the system and given explicitly by \cite{potts2020magnon}
\begin{equation}
    \Sigma[\omega] = i\vert g_{\rm mb} \vert^2 ( \Xi[\omega] - \Xi^*[-\omega]).
    \label{Eqn:06}
\end{equation}
Here, $g_{\rm mb} = g^0_{\rm mb}\langle \hat{m}\rangle$ is the cavity-enhanced magnomechanical coupling rate, $\vert \langle \hat{m} \rangle\vert^2$ is the steady-state magnon population, and $\Xi^{-1}[\omega] = \chi_{\rm m}^{-1}[\omega] + g_{\rm am}^2\chi_{\rm a}[\omega]$. The magnon and cavity susceptibilities are given by $\chi_{\textrm{m}}[\omega] = [-i(\Delta_{\textrm{m}}+\omega) + \gamma_{\textrm{m}}/2]$ and $\chi_{\textrm{a}}[\omega] = [-i(\Delta_{\textrm{a}}+\omega) + \kappa/2]$, respectively. In the weak coupling limit, $g_{\rm mb} \ll \{ \kappa, \gamma_{\rm m} \}$ the real and imaginary part of the self-energy describe a shift in the mechanical frequency and the magnomechanical damping rate, respectively. 

\begin{figure}[t]
\includegraphics{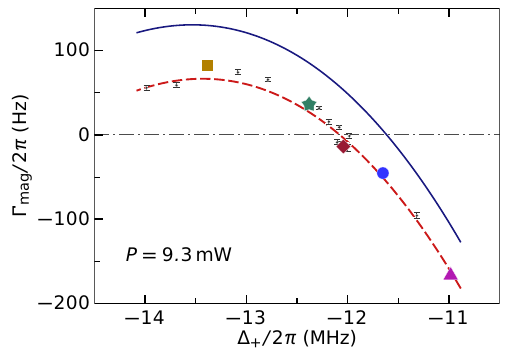}
\caption{Magnomechanical damping rate as a function of the detuning at a constant power of 9.3 mW. Blue solid curve is given by the linear theory given by $\Gamma_{\rm mag} = 2 \Im\Sigma[\omega]$. The red dashed curve is the shifted linear theory given by Eq.~\eqref{Eqn:07}. Fits were performed on all powers simultaneously (not shown), giving $g_{\rm mb}^0/2\pi = 4.56$ mHz and $\alpha/2\pi = -1.24$ pHz. Error bars are the statistical error over one hundred independent experimental runs. The colored data points indicate the data within the corresponding curve in Fig.~\ref{Fig:04}.}
\label{Fig:05}
\end{figure}

For every experimental run, a simultaneous sweep of the normal mode was fit to extract all parameters regarding the cavity and magnon, leaving only a single fit parameter, the single-magnon magnomechanical coupling rate $g_{\rm mb}^0$. As shown in Fig.~\ref{Fig:05}, using the linear theory predicted from Eq.~\eqref{Eqn:06} results in a discrepancy between the theory and measured data. For all detunings and powers, the measured magnomechanical damping rate was smaller than the rate predicted. Furthermore, the difference between the measured data and the theory increases approximately linearly with drive power (see Supplementary Material \cite{SupplimentaryInfo}). We suspect this difference is a result of two effects: first, the intrinsic Kerr nonlinearity of YIG results in a modification of the self-energy, and secondly, we suspect that the phonon mode is coupling to weakly driven higher-order magnon modes \cite{BittencourtDamping2022}.

We add a correction to the magnomechanical damping rate that phenomenologically models the phonon mode interacting with weakly driven higher-order magnon modes \cite{BittencourtDamping2022}, such that
\begin{equation}
    \Gamma_{\rm mag} = 2 \Im\Sigma[\omega] + \alpha \vert \langle \hat{m} \rangle \vert^2.
    \label{Eqn:07}
\end{equation}
The term $\alpha \vert\langle \hat{m} \rangle \vert^2$ is proportional to the drive power and we find good agreement between the measured data and the modified magnomechanical damping rate. It should be noted that we perform the fit to all powers simultaneously; we only show one power as an example. From the fit, we extract a single-magnon magnomechanical coupling rate $g_{\rm mb}^0/2\pi = 4.56\,$mHz, which is in excellent agreement with previous experiments \cite{zhang2016cavity,potts2021dynamical}, and the linear correction is given by $\alpha/2\pi = -1.24\,$pHz.

Cavity magnomechanics has recently seen considerable experimental and theoretical attention \cite{fan2022microwave,fan2022microwave,li2021entangling,li2019entangling,li2018magnon,li2021squeezing,ding2022magnon,asjad2022magnon,li2019squeezed,zhang2021generation,sarma2021cavity,li2020phase,kong2019magnetically,li2021quantum,hatanaka2022chip}, with experimental observations such as MMIT \cite{zhang2016cavity}, dynamical backaction \cite{potts2021dynamical}, and Kerr bistability \cite{shen2022mechanical}. Here, we demonstrate the ability to tune smoothly between magnomechanical damping and antidamping. Moreover, we eliminate dynamical backaction from our magnomechanical measurement, a feature enabled by the two-phonon triple resonance magnomechanical system. We also observe a deviation in the behavior of the magnomechanical decay rate from the previous magnomechanical backaction theory. We attribute this difference to weakly driven magnon modes and the intrinsic Kerr nonlinearity of YIG, which can be taken into account via a power-dependent correction \cite{BittencourtDamping2022}.

Looking forward, the elimination of dynamical backaction could enable, for example, the future observation of magnon-induced quantum backaction or the implementation of magnomechanical-based primary thermometry \cite{potts2020magnon}. Furthermore, the ability to rapidly and accurately modulate the magnomechanical damping rate provides the foundation for future experiments exploring, for example, macroscopic entanglement swapping \cite{kotler2021direct}. The large effective mechanical mass of the YIG sphere --- compared to conventional optomechanical experiments \cite{kotler2021direct,rodrigues2019coupling} --- makes cavity magnomechanics an exciting candidate for testing exotic theories, such as gravitational decoherence in quantum mechanics \cite{penrose1996gravity,diosi1987universal,kanari2021can,gely2021superconducting,Marius_Schrodinger_2022}.

%Moreover, with additional carefully tuned drive tones, it may be possible in the future to demonstrate a measurement that evades quantum backaction. 

\begin{acknowledgments}
The authors acknowledge helpful contributions from E. Varga and that the land on which this work was performed is in Treaty Six Territory, the traditional territories of many First Nations, Métis, and Inuit in Alberta. V.A.S.V.B. thanks S. Scharma for the useful discussions. This work was supported by the University of Alberta; the Natural Sciences and Engineering Research Council, Canada (Grant Nos.~RGPIN-2016-04523, RGPIN-2022-03078, and CREATE-495446-17); the Alberta Quantum Major Innovation Fund; and the Government of Canada through the NRC Quantum Sensors Program. V.A.S.V. Bittencourt and S. Viola Kusminskiy acknowledge financial support from the Max Planck Society and from the Deutsche Forschungsgemeinschaft (DFG, German Research Foundation) through Project-ID 429529648–TRR 306 QuCoLiMa (“Quantum Cooperativity of Light and Matter”).
\end{acknowledgments}

\bibliography{apssamp}% Produces the bibliography via BibTeX.

\providecommand{\noopsort}[1]{}\providecommand{\singleletter}[1]{#1}%
\begin{thebibliography}{69}%
\makeatletter
\providecommand \@ifxundefined [1]{%
 \@ifx{#1\undefined}
}%
\providecommand \@ifnum [1]{%
 \ifnum #1\expandafter \@firstoftwo
 \else \expandafter \@secondoftwo
 \fi
}%
\providecommand \@ifx [1]{%
 \ifx #1\expandafter \@firstoftwo
 \else \expandafter \@secondoftwo
 \fi
}%
\providecommand \natexlab [1]{#1}%
\providecommand \enquote  [1]{``#1''}%
\providecommand \bibnamefont  [1]{#1}%
\providecommand \bibfnamefont [1]{#1}%
\providecommand \citenamefont [1]{#1}%
\providecommand \href@noop [0]{\@secondoftwo}%
\providecommand \href [0]{\begingroup \@sanitize@url \@href}%
\providecommand \@href[1]{\@@startlink{#1}\@@href}%
\providecommand \@@href[1]{\endgroup#1\@@endlink}%
\providecommand \@sanitize@url [0]{\catcode `\\12\catcode `\$12\catcode
  `\&12\catcode `\#12\catcode `\^12\catcode `\_12\catcode `\%12\relax}%
\providecommand \@@startlink[1]{}%
\providecommand \@@endlink[0]{}%
\providecommand \url  [0]{\begingroup\@sanitize@url \@url }%
\providecommand \@url [1]{\endgroup\@href {#1}{\urlprefix }}%
\providecommand \urlprefix  [0]{URL }%
\providecommand \Eprint [0]{\href }%
\providecommand \doibase [0]{https://doi.org/}%
\providecommand \selectlanguage [0]{\@gobble}%
\providecommand \bibinfo  [0]{\@secondoftwo}%
\providecommand \bibfield  [0]{\@secondoftwo}%
\providecommand \translation [1]{[#1]}%
\providecommand \BibitemOpen [0]{}%
\providecommand \bibitemStop [0]{}%
\providecommand \bibitemNoStop [0]{.\EOS\space}%
\providecommand \EOS [0]{\spacefactor3000\relax}%
\providecommand \BibitemShut  [1]{\csname bibitem#1\endcsname}%
\let\auto@bib@innerbib\@empty
%</preamble>
\bibitem [{\citenamefont {Ashkin}(1980)}]{ashkin1980applications}%
  \BibitemOpen
  \bibfield  {author} {\bibinfo {author} {\bibfnamefont {A.}~\bibnamefont
  {Ashkin}},\ }\bibfield  {title} {\bibinfo {title} {Applications of laser
  radiation pressure},\ }\href@noop {} {\bibfield  {journal} {\bibinfo
  {journal} {Science}\ }\textbf {\bibinfo {volume} {210}},\ \bibinfo {pages}
  {1081} (\bibinfo {year} {1980})}\BibitemShut {NoStop}%
\bibitem [{\citenamefont {Cohadon}\ \emph {et~al.}(1999)\citenamefont
  {Cohadon}, \citenamefont {Heidmann},\ and\ \citenamefont
  {Pinard}}]{cohadon1999cooling}%
  \BibitemOpen
  \bibfield  {author} {\bibinfo {author} {\bibfnamefont {P.-F.}\ \bibnamefont
  {Cohadon}}, \bibinfo {author} {\bibfnamefont {A.}~\bibnamefont {Heidmann}},\
  and\ \bibinfo {author} {\bibfnamefont {M.}~\bibnamefont {Pinard}},\
  }\bibfield  {title} {\bibinfo {title} {Cooling of a mirror by radiation
  pressure},\ }\href@noop {} {\bibfield  {journal} {\bibinfo  {journal} {Phys.
  Rev. Lett.}\ }\textbf {\bibinfo {volume} {83}},\ \bibinfo {pages} {3174}
  (\bibinfo {year} {1999})}\BibitemShut {NoStop}%
\bibitem [{\citenamefont {Metzger}\ and\ \citenamefont
  {Karrai}(2004)}]{metzger2004cavity}%
  \BibitemOpen
  \bibfield  {author} {\bibinfo {author} {\bibfnamefont {C.~H.}\ \bibnamefont
  {Metzger}}\ and\ \bibinfo {author} {\bibfnamefont {K.}~\bibnamefont
  {Karrai}},\ }\bibfield  {title} {\bibinfo {title} {Cavity cooling of a
  microlever},\ }\href@noop {} {\bibfield  {journal} {\bibinfo  {journal}
  {Nature}\ }\textbf {\bibinfo {volume} {432}},\ \bibinfo {pages} {1002}
  (\bibinfo {year} {2004})}\BibitemShut {NoStop}%
\bibitem [{\citenamefont {Aspelmeyer}\ \emph {et~al.}(2014)\citenamefont
  {Aspelmeyer}, \citenamefont {Kippenberg},\ and\ \citenamefont
  {Marquardt}}]{aspelmeyer2014cavity}%
  \BibitemOpen
  \bibfield  {author} {\bibinfo {author} {\bibfnamefont {M.}~\bibnamefont
  {Aspelmeyer}}, \bibinfo {author} {\bibfnamefont {T.~J.}\ \bibnamefont
  {Kippenberg}},\ and\ \bibinfo {author} {\bibfnamefont {F.}~\bibnamefont
  {Marquardt}},\ }\bibfield  {title} {\bibinfo {title} {Cavity optomechanics},\
  }\href@noop {} {\bibfield  {journal} {\bibinfo  {journal} {Rev. Mod. Phys.}\
  }\textbf {\bibinfo {volume} {86}},\ \bibinfo {pages} {1391} (\bibinfo {year}
  {2014})}\BibitemShut {NoStop}%
\bibitem [{\citenamefont {Chan}\ \emph {et~al.}(2011)\citenamefont {Chan},
  \citenamefont {Alegre}, \citenamefont {Safavi-Naeini}, \citenamefont {Hill},
  \citenamefont {Krause}, \citenamefont {Gr{\"o}blacher}, \citenamefont
  {Aspelmeyer},\ and\ \citenamefont {Painter}}]{chan2011laser}%
  \BibitemOpen
  \bibfield  {author} {\bibinfo {author} {\bibfnamefont {J.}~\bibnamefont
  {Chan}}, \bibinfo {author} {\bibfnamefont {T.}~\bibnamefont {Alegre}},
  \bibinfo {author} {\bibfnamefont {A.~H.}\ \bibnamefont {Safavi-Naeini}},
  \bibinfo {author} {\bibfnamefont {J.~T.}\ \bibnamefont {Hill}}, \bibinfo
  {author} {\bibfnamefont {A.}~\bibnamefont {Krause}}, \bibinfo {author}
  {\bibfnamefont {S.}~\bibnamefont {Gr{\"o}blacher}}, \bibinfo {author}
  {\bibfnamefont {M.}~\bibnamefont {Aspelmeyer}},\ and\ \bibinfo {author}
  {\bibfnamefont {O.}~\bibnamefont {Painter}},\ }\bibfield  {title} {\bibinfo
  {title} {Laser cooling of a nanomechanical oscillator into its quantum ground
  state},\ }\href@noop {} {\bibfield  {journal} {\bibinfo  {journal} {Nature}\
  }\textbf {\bibinfo {volume} {478}},\ \bibinfo {pages} {89} (\bibinfo {year}
  {2011})}\BibitemShut {NoStop}%
\bibitem [{\citenamefont {Teufel}\ \emph {et~al.}(2011)\citenamefont {Teufel},
  \citenamefont {Donner}, \citenamefont {Li}, \citenamefont {Harlow},
  \citenamefont {Allman}, \citenamefont {Cicak}, \citenamefont {Sirois},
  \citenamefont {Whittaker}, \citenamefont {Lehnert},\ and\ \citenamefont
  {Simmonds}}]{teufel2011sideband}%
  \BibitemOpen
  \bibfield  {author} {\bibinfo {author} {\bibfnamefont {J.~D.}\ \bibnamefont
  {Teufel}}, \bibinfo {author} {\bibfnamefont {T.}~\bibnamefont {Donner}},
  \bibinfo {author} {\bibfnamefont {D.}~\bibnamefont {Li}}, \bibinfo {author}
  {\bibfnamefont {J.~W.}\ \bibnamefont {Harlow}}, \bibinfo {author}
  {\bibfnamefont {M.}~\bibnamefont {Allman}}, \bibinfo {author} {\bibfnamefont
  {K.}~\bibnamefont {Cicak}}, \bibinfo {author} {\bibfnamefont {A.~J.}\
  \bibnamefont {Sirois}}, \bibinfo {author} {\bibfnamefont {J.~D.}\
  \bibnamefont {Whittaker}}, \bibinfo {author} {\bibfnamefont {K.~W.}\
  \bibnamefont {Lehnert}},\ and\ \bibinfo {author} {\bibfnamefont {R.~W.}\
  \bibnamefont {Simmonds}},\ }\bibfield  {title} {\bibinfo {title} {Sideband
  cooling of micromechanical motion to the quantum ground state},\ }\href@noop
  {} {\bibfield  {journal} {\bibinfo  {journal} {Nature}\ }\textbf {\bibinfo
  {volume} {475}},\ \bibinfo {pages} {359} (\bibinfo {year}
  {2011})}\BibitemShut {NoStop}%
\bibitem [{\citenamefont {Chumak}\ \emph {et~al.}(2021)\citenamefont {Chumak},
  \citenamefont {Kabos}, \citenamefont {Wu}, \citenamefont {Abert},
  \citenamefont {Adelmann}, \citenamefont {Adeyeye}, \citenamefont
  {{\AA}kerman}, \citenamefont {Aliev}, \citenamefont {Anane}, \citenamefont
  {Awad} \emph {et~al.}}]{chumak2021roadmap}%
  \BibitemOpen
  \bibfield  {author} {\bibinfo {author} {\bibfnamefont {A.}~\bibnamefont
  {Chumak}}, \bibinfo {author} {\bibfnamefont {P.}~\bibnamefont {Kabos}},
  \bibinfo {author} {\bibfnamefont {M.}~\bibnamefont {Wu}}, \bibinfo {author}
  {\bibfnamefont {C.}~\bibnamefont {Abert}}, \bibinfo {author} {\bibfnamefont
  {C.}~\bibnamefont {Adelmann}}, \bibinfo {author} {\bibfnamefont
  {A.}~\bibnamefont {Adeyeye}}, \bibinfo {author} {\bibfnamefont
  {J.}~\bibnamefont {{\AA}kerman}}, \bibinfo {author} {\bibfnamefont
  {F.}~\bibnamefont {Aliev}}, \bibinfo {author} {\bibfnamefont
  {A.}~\bibnamefont {Anane}}, \bibinfo {author} {\bibfnamefont
  {A.}~\bibnamefont {Awad}}, \emph {et~al.},\ }\bibfield  {title} {\bibinfo
  {title} {Roadmap on spin-wave computing concepts},\ }\href@noop {} {\bibfield
   {journal} {\bibinfo  {journal} {IEEE Trans. Quantum Eng.}\ } (\bibinfo
  {year} {2021})}\BibitemShut {NoStop}%
\bibitem [{\citenamefont {Rameshti}\ \emph {et~al.}(2022)\citenamefont
  {Rameshti}, \citenamefont {Viola~Kusminskiy}, \citenamefont {Haigh},
  \citenamefont {Usami}, \citenamefont {Lachance-Quirion}, \citenamefont
  {Nakamura}, \citenamefont {Hu}, \citenamefont {Tang}, \citenamefont {Bauer},\
  and\ \citenamefont {Blanter}}]{rameshti2022cavity}%
  \BibitemOpen
  \bibfield  {author} {\bibinfo {author} {\bibfnamefont {B.~Z.}\ \bibnamefont
  {Rameshti}}, \bibinfo {author} {\bibfnamefont {S.}~\bibnamefont
  {Viola~Kusminskiy}}, \bibinfo {author} {\bibfnamefont {J.~A.}\ \bibnamefont
  {Haigh}}, \bibinfo {author} {\bibfnamefont {K.}~\bibnamefont {Usami}},
  \bibinfo {author} {\bibfnamefont {D.}~\bibnamefont {Lachance-Quirion}},
  \bibinfo {author} {\bibfnamefont {Y.}~\bibnamefont {Nakamura}}, \bibinfo
  {author} {\bibfnamefont {C.-M.}\ \bibnamefont {Hu}}, \bibinfo {author}
  {\bibfnamefont {H.~X.}\ \bibnamefont {Tang}}, \bibinfo {author}
  {\bibfnamefont {G.~E.}\ \bibnamefont {Bauer}},\ and\ \bibinfo {author}
  {\bibfnamefont {Y.~M.}\ \bibnamefont {Blanter}},\ }\bibfield  {title}
  {\bibinfo {title} {Cavity magnonics},\ }\href@noop {} {\bibfield  {journal}
  {\bibinfo  {journal} {Phys. Rep.}\ }\textbf {\bibinfo {volume} {979}},\
  \bibinfo {pages} {1} (\bibinfo {year} {2022})}\BibitemShut {NoStop}%
\bibitem [{\citenamefont {Lachance-Quirion}\ \emph {et~al.}(2019)\citenamefont
  {Lachance-Quirion}, \citenamefont {Tabuchi}, \citenamefont {Gloppe},
  \citenamefont {Usami},\ and\ \citenamefont {Nakamura}}]{lachance2019hybrid}%
  \BibitemOpen
  \bibfield  {author} {\bibinfo {author} {\bibfnamefont {D.}~\bibnamefont
  {Lachance-Quirion}}, \bibinfo {author} {\bibfnamefont {Y.}~\bibnamefont
  {Tabuchi}}, \bibinfo {author} {\bibfnamefont {A.}~\bibnamefont {Gloppe}},
  \bibinfo {author} {\bibfnamefont {K.}~\bibnamefont {Usami}},\ and\ \bibinfo
  {author} {\bibfnamefont {Y.}~\bibnamefont {Nakamura}},\ }\bibfield  {title}
  {\bibinfo {title} {Hybrid quantum systems based on magnonics},\ }\href@noop
  {} {\bibfield  {journal} {\bibinfo  {journal} {Appl. Phys. Express}\ }\textbf
  {\bibinfo {volume} {12}},\ \bibinfo {pages} {070101} (\bibinfo {year}
  {2019})}\BibitemShut {NoStop}%
\bibitem [{\citenamefont {Zhang}\ \emph {et~al.}(2016)\citenamefont {Zhang},
  \citenamefont {Zou}, \citenamefont {Jiang},\ and\ \citenamefont
  {Tang}}]{zhang2016cavity}%
  \BibitemOpen
  \bibfield  {author} {\bibinfo {author} {\bibfnamefont {X.}~\bibnamefont
  {Zhang}}, \bibinfo {author} {\bibfnamefont {C.-L.}\ \bibnamefont {Zou}},
  \bibinfo {author} {\bibfnamefont {L.}~\bibnamefont {Jiang}},\ and\ \bibinfo
  {author} {\bibfnamefont {H.~X.}\ \bibnamefont {Tang}},\ }\bibfield  {title}
  {\bibinfo {title} {Cavity magnomechanics},\ }\href@noop {} {\bibfield
  {journal} {\bibinfo  {journal} {Sci. Adv.}\ }\textbf {\bibinfo {volume}
  {2}},\ \bibinfo {pages} {e1501286} (\bibinfo {year} {2016})}\BibitemShut
  {NoStop}%
\bibitem [{\citenamefont {Potts}\ \emph {et~al.}(2021)\citenamefont {Potts},
  \citenamefont {Varga}, \citenamefont {Bittencourt}, \citenamefont
  {Viola~Kusminskiy},\ and\ \citenamefont {Davis}}]{potts2021dynamical}%
  \BibitemOpen
  \bibfield  {author} {\bibinfo {author} {\bibfnamefont {C.~A.}\ \bibnamefont
  {Potts}}, \bibinfo {author} {\bibfnamefont {E.}~\bibnamefont {Varga}},
  \bibinfo {author} {\bibfnamefont {V.~A. S.~V.}\ \bibnamefont {Bittencourt}},
  \bibinfo {author} {\bibfnamefont {S.}~\bibnamefont {Viola~Kusminskiy}},\ and\
  \bibinfo {author} {\bibfnamefont {J.~P.}\ \bibnamefont {Davis}},\ }\bibfield
  {title} {\bibinfo {title} {Dynamical backaction magnomechanics},\ }\href
  {https://doi.org/10.1103/PhysRevX.11.031053} {\bibfield  {journal} {\bibinfo
  {journal} {Phys. Rev. X}\ }\textbf {\bibinfo {volume} {11}},\ \bibinfo
  {pages} {031053} (\bibinfo {year} {2021})}\BibitemShut {NoStop}%
\bibitem [{\citenamefont {Shen}\ \emph {et~al.}(2022)\citenamefont {Shen},
  \citenamefont {Li}, \citenamefont {Fan}, \citenamefont {Wang},\ and\
  \citenamefont {You}}]{shen2022mechanical}%
  \BibitemOpen
  \bibfield  {author} {\bibinfo {author} {\bibfnamefont {R.-C.}\ \bibnamefont
  {Shen}}, \bibinfo {author} {\bibfnamefont {J.}~\bibnamefont {Li}}, \bibinfo
  {author} {\bibfnamefont {Z.-Y.}\ \bibnamefont {Fan}}, \bibinfo {author}
  {\bibfnamefont {Y.-P.}\ \bibnamefont {Wang}},\ and\ \bibinfo {author}
  {\bibfnamefont {J.~Q.}\ \bibnamefont {You}},\ }\bibfield  {title} {\bibinfo
  {title} {Mechanical bistability in kerr-modified cavity magnomechanics},\
  }\href@noop {} {\bibfield  {journal} {\bibinfo  {journal} {Phys. Rev. Lett.}\
  }\textbf {\bibinfo {volume} {129}},\ \bibinfo {pages} {123601} (\bibinfo
  {year} {2022})}\BibitemShut {NoStop}%
\bibitem [{Fer(2022)}]{FerriSphere}%
  \BibitemOpen
  \href@noop {} {}\bibinfo {howpublished} {See
  \url{http://www.ferrisphere.com/} for information on purchasing YIG spheres.}
  (\bibinfo {year} {2022}),\ \bibinfo {note} {accessed: 21-10-2022}\BibitemShut
  {NoStop}%
\bibitem [{\citenamefont {Huebl}\ \emph {et~al.}(2013)\citenamefont {Huebl},
  \citenamefont {Zollitsch}, \citenamefont {Lotze}, \citenamefont {Hocke},
  \citenamefont {Greifenstein}, \citenamefont {Marx}, \citenamefont {Gross},\
  and\ \citenamefont {Goennenwein}}]{huebl2013high}%
  \BibitemOpen
  \bibfield  {author} {\bibinfo {author} {\bibfnamefont {H.}~\bibnamefont
  {Huebl}}, \bibinfo {author} {\bibfnamefont {C.~W.}\ \bibnamefont
  {Zollitsch}}, \bibinfo {author} {\bibfnamefont {J.}~\bibnamefont {Lotze}},
  \bibinfo {author} {\bibfnamefont {F.}~\bibnamefont {Hocke}}, \bibinfo
  {author} {\bibfnamefont {M.}~\bibnamefont {Greifenstein}}, \bibinfo {author}
  {\bibfnamefont {A.}~\bibnamefont {Marx}}, \bibinfo {author} {\bibfnamefont
  {R.}~\bibnamefont {Gross}},\ and\ \bibinfo {author} {\bibfnamefont
  {S.~T.~B.}\ \bibnamefont {Goennenwein}},\ }\bibfield  {title} {\bibinfo
  {title} {High cooperativity in coupled microwave resonator ferrimagnetic
  insulator hybrids},\ }\href@noop {} {\bibfield  {journal} {\bibinfo
  {journal} {Phys. Rev. Lett.}\ }\textbf {\bibinfo {volume} {111}},\ \bibinfo
  {pages} {127003} (\bibinfo {year} {2013})}\BibitemShut {NoStop}%
\bibitem [{\citenamefont {Zhang}\ \emph {et~al.}(2014)\citenamefont {Zhang},
  \citenamefont {Zou}, \citenamefont {Jiang},\ and\ \citenamefont
  {Tang}}]{zhang2014strongly}%
  \BibitemOpen
  \bibfield  {author} {\bibinfo {author} {\bibfnamefont {X.}~\bibnamefont
  {Zhang}}, \bibinfo {author} {\bibfnamefont {C.-L.}\ \bibnamefont {Zou}},
  \bibinfo {author} {\bibfnamefont {L.}~\bibnamefont {Jiang}},\ and\ \bibinfo
  {author} {\bibfnamefont {H.~X.}\ \bibnamefont {Tang}},\ }\bibfield  {title}
  {\bibinfo {title} {Strongly coupled magnons and cavity microwave photons},\
  }\href@noop {} {\bibfield  {journal} {\bibinfo  {journal} {Phys. Rev. Lett.}\
  }\textbf {\bibinfo {volume} {113}},\ \bibinfo {pages} {156401} (\bibinfo
  {year} {2014})}\BibitemShut {NoStop}%
\bibitem [{\citenamefont {Tabuchi}\ \emph {et~al.}(2015)\citenamefont
  {Tabuchi}, \citenamefont {Ishino}, \citenamefont {Noguchi}, \citenamefont
  {Ishikawa}, \citenamefont {Yamazaki}, \citenamefont {Usami},\ and\
  \citenamefont {Nakamura}}]{tabuchi2015coherent}%
  \BibitemOpen
  \bibfield  {author} {\bibinfo {author} {\bibfnamefont {Y.}~\bibnamefont
  {Tabuchi}}, \bibinfo {author} {\bibfnamefont {S.}~\bibnamefont {Ishino}},
  \bibinfo {author} {\bibfnamefont {A.}~\bibnamefont {Noguchi}}, \bibinfo
  {author} {\bibfnamefont {T.}~\bibnamefont {Ishikawa}}, \bibinfo {author}
  {\bibfnamefont {R.}~\bibnamefont {Yamazaki}}, \bibinfo {author}
  {\bibfnamefont {K.}~\bibnamefont {Usami}},\ and\ \bibinfo {author}
  {\bibfnamefont {Y.}~\bibnamefont {Nakamura}},\ }\bibfield  {title} {\bibinfo
  {title} {Coherent coupling between a ferromagnetic magnon and a
  superconducting qubit},\ }\href@noop {} {\bibfield  {journal} {\bibinfo
  {journal} {Science}\ }\textbf {\bibinfo {volume} {349}},\ \bibinfo {pages}
  {405} (\bibinfo {year} {2015})}\BibitemShut {NoStop}%
\bibitem [{\citenamefont {Bourcin}\ \emph {et~al.}(2022)\citenamefont
  {Bourcin}, \citenamefont {Bourhill}, \citenamefont {Vlaminck},\ and\
  \citenamefont {Castel}}]{bourcin2022strong}%
  \BibitemOpen
  \bibfield  {author} {\bibinfo {author} {\bibfnamefont {G.}~\bibnamefont
  {Bourcin}}, \bibinfo {author} {\bibfnamefont {J.}~\bibnamefont {Bourhill}},
  \bibinfo {author} {\bibfnamefont {V.}~\bibnamefont {Vlaminck}},\ and\
  \bibinfo {author} {\bibfnamefont {V.}~\bibnamefont {Castel}},\ }\bibfield
  {title} {\bibinfo {title} {Strong to ultra-strong coherent coupling
  measurements in a yig/cavity system at room temperature},\ }\href@noop {}
  {\bibfield  {journal} {\bibinfo  {journal} {arXiv:2209.14643}\ } (\bibinfo
  {year} {2022})}\BibitemShut {NoStop}%
\bibitem [{\citenamefont {Potts}\ and\ \citenamefont
  {Davis}(2020)}]{potts2020strong}%
  \BibitemOpen
  \bibfield  {author} {\bibinfo {author} {\bibfnamefont {C.~A.}\ \bibnamefont
  {Potts}}\ and\ \bibinfo {author} {\bibfnamefont {J.~P.}\ \bibnamefont
  {Davis}},\ }\bibfield  {title} {\bibinfo {title} {Strong magnon--photon
  coupling within a tunable cryogenic microwave cavity},\ }\href@noop {}
  {\bibfield  {journal} {\bibinfo  {journal} {Appl. Phys. Lett.}\ }\textbf
  {\bibinfo {volume} {116}},\ \bibinfo {pages} {263503} (\bibinfo {year}
  {2020})}\BibitemShut {NoStop}%
\bibitem [{\citenamefont {Wang}\ \emph {et~al.}(2019)\citenamefont {Wang},
  \citenamefont {Rao}, \citenamefont {Yang}, \citenamefont {Xu}, \citenamefont
  {Gui}, \citenamefont {Yao}, \citenamefont {You},\ and\ \citenamefont
  {Hu}}]{wang2019nonreciprocity}%
  \BibitemOpen
  \bibfield  {author} {\bibinfo {author} {\bibfnamefont {Y.-P.}\ \bibnamefont
  {Wang}}, \bibinfo {author} {\bibfnamefont {J.~W.}\ \bibnamefont {Rao}},
  \bibinfo {author} {\bibfnamefont {Y.}~\bibnamefont {Yang}}, \bibinfo {author}
  {\bibfnamefont {P.-C.}\ \bibnamefont {Xu}}, \bibinfo {author} {\bibfnamefont
  {Y.~S.}\ \bibnamefont {Gui}}, \bibinfo {author} {\bibfnamefont {B.~M.}\
  \bibnamefont {Yao}}, \bibinfo {author} {\bibfnamefont {J.~Q.}\ \bibnamefont
  {You}},\ and\ \bibinfo {author} {\bibfnamefont {C.-M.}\ \bibnamefont {Hu}},\
  }\bibfield  {title} {\bibinfo {title} {Nonreciprocity and unidirectional
  invisibility in cavity magnonics},\ }\href@noop {} {\bibfield  {journal}
  {\bibinfo  {journal} {Phys. Rev. Lett.}\ }\textbf {\bibinfo {volume} {123}},\
  \bibinfo {pages} {127202} (\bibinfo {year} {2019})}\BibitemShut {NoStop}%
\bibitem [{\citenamefont {Harder}\ \emph {et~al.}(2018)\citenamefont {Harder},
  \citenamefont {Yang}, \citenamefont {Yao}, \citenamefont {Yu}, \citenamefont
  {Rao}, \citenamefont {Gui}, \citenamefont {Stamps},\ and\ \citenamefont
  {Hu}}]{harder2018level}%
  \BibitemOpen
  \bibfield  {author} {\bibinfo {author} {\bibfnamefont {M.}~\bibnamefont
  {Harder}}, \bibinfo {author} {\bibfnamefont {Y.}~\bibnamefont {Yang}},
  \bibinfo {author} {\bibfnamefont {B.~M.}\ \bibnamefont {Yao}}, \bibinfo
  {author} {\bibfnamefont {C.~H.}\ \bibnamefont {Yu}}, \bibinfo {author}
  {\bibfnamefont {J.~W.}\ \bibnamefont {Rao}}, \bibinfo {author} {\bibfnamefont
  {Y.~S.}\ \bibnamefont {Gui}}, \bibinfo {author} {\bibfnamefont {R.~L.}\
  \bibnamefont {Stamps}},\ and\ \bibinfo {author} {\bibfnamefont {C.-M.}\
  \bibnamefont {Hu}},\ }\bibfield  {title} {\bibinfo {title} {Level attraction
  due to dissipative magnon-photon coupling},\ }\href@noop {} {\bibfield
  {journal} {\bibinfo  {journal} {Phys. Rev. Lett.}\ }\textbf {\bibinfo
  {volume} {121}},\ \bibinfo {pages} {137203} (\bibinfo {year}
  {2018})}\BibitemShut {NoStop}%
\bibitem [{\citenamefont {Zhang}\ \emph {et~al.}(2015)\citenamefont {Zhang},
  \citenamefont {Zou}, \citenamefont {Zhu}, \citenamefont {Marquardt},
  \citenamefont {Jiang},\ and\ \citenamefont {Tang}}]{zhang2015magnon}%
  \BibitemOpen
  \bibfield  {author} {\bibinfo {author} {\bibfnamefont {X.}~\bibnamefont
  {Zhang}}, \bibinfo {author} {\bibfnamefont {C.-L.}\ \bibnamefont {Zou}},
  \bibinfo {author} {\bibfnamefont {N.}~\bibnamefont {Zhu}}, \bibinfo {author}
  {\bibfnamefont {F.}~\bibnamefont {Marquardt}}, \bibinfo {author}
  {\bibfnamefont {L.}~\bibnamefont {Jiang}},\ and\ \bibinfo {author}
  {\bibfnamefont {H.~X.}\ \bibnamefont {Tang}},\ }\bibfield  {title} {\bibinfo
  {title} {Magnon dark modes and gradient memory},\ }\href@noop {} {\bibfield
  {journal} {\bibinfo  {journal} {Nat. Commun.}\ }\textbf {\bibinfo {volume}
  {6}},\ \bibinfo {pages} {1} (\bibinfo {year} {2015})}\BibitemShut {NoStop}%
\bibitem [{\citenamefont {Morris}\ \emph {et~al.}(2017)\citenamefont {Morris},
  \citenamefont {Van~Loo}, \citenamefont {Kosen},\ and\ \citenamefont
  {Karenowska}}]{morris2017strong}%
  \BibitemOpen
  \bibfield  {author} {\bibinfo {author} {\bibfnamefont {R.}~\bibnamefont
  {Morris}}, \bibinfo {author} {\bibfnamefont {A.}~\bibnamefont {Van~Loo}},
  \bibinfo {author} {\bibfnamefont {S.}~\bibnamefont {Kosen}},\ and\ \bibinfo
  {author} {\bibfnamefont {A.}~\bibnamefont {Karenowska}},\ }\bibfield  {title}
  {\bibinfo {title} {Strong coupling of magnons in a yig sphere to photons in a
  planar superconducting resonator in the quantum limit},\ }\href@noop {}
  {\bibfield  {journal} {\bibinfo  {journal} {Sci. Rep.}\ }\textbf {\bibinfo
  {volume} {7}},\ \bibinfo {pages} {1} (\bibinfo {year} {2017})}\BibitemShut
  {NoStop}%
\bibitem [{\citenamefont {Streib}\ \emph {et~al.}(2019)\citenamefont {Streib},
  \citenamefont {Vidal-Silva}, \citenamefont {Shen},\ and\ \citenamefont
  {Bauer}}]{streib2019magnon}%
  \BibitemOpen
  \bibfield  {author} {\bibinfo {author} {\bibfnamefont {S.}~\bibnamefont
  {Streib}}, \bibinfo {author} {\bibfnamefont {N.}~\bibnamefont {Vidal-Silva}},
  \bibinfo {author} {\bibfnamefont {K.}~\bibnamefont {Shen}},\ and\ \bibinfo
  {author} {\bibfnamefont {G.~E.~W.}\ \bibnamefont {Bauer}},\ }\bibfield
  {title} {\bibinfo {title} {Magnon-phonon interactions in magnetic
  insulators},\ }\href@noop {} {\bibfield  {journal} {\bibinfo  {journal}
  {Phys. Rev. B}\ }\textbf {\bibinfo {volume} {99}},\ \bibinfo {pages} {184442}
  (\bibinfo {year} {2019})}\BibitemShut {NoStop}%
\bibitem [{\citenamefont {Walker}(1958)}]{walker1958resonant}%
  \BibitemOpen
  \bibfield  {author} {\bibinfo {author} {\bibfnamefont {L.}~\bibnamefont
  {Walker}},\ }\bibfield  {title} {\bibinfo {title} {Resonant modes of
  ferromagnetic spheroids},\ }\href@noop {} {\bibfield  {journal} {\bibinfo
  {journal} {J. Appl. Phys.}\ }\textbf {\bibinfo {volume} {29}},\ \bibinfo
  {pages} {318} (\bibinfo {year} {1958})}\BibitemShut {NoStop}%
\bibitem [{\citenamefont {Fletcher}\ and\ \citenamefont
  {Bell}(1959)}]{fletcher1959ferrimagnetic}%
  \BibitemOpen
  \bibfield  {author} {\bibinfo {author} {\bibfnamefont {P.}~\bibnamefont
  {Fletcher}}\ and\ \bibinfo {author} {\bibfnamefont {R.}~\bibnamefont
  {Bell}},\ }\bibfield  {title} {\bibinfo {title} {Ferrimagnetic resonance
  modes in spheres},\ }\href@noop {} {\bibfield  {journal} {\bibinfo  {journal}
  {J. Appl. Phys.}\ }\textbf {\bibinfo {volume} {30}},\ \bibinfo {pages} {687}
  (\bibinfo {year} {1959})}\BibitemShut {NoStop}%
\bibitem [{\citenamefont {Schlömann}(1960)}]{Schlomann1960generation}%
  \BibitemOpen
  \bibfield  {author} {\bibinfo {author} {\bibfnamefont {E.}~\bibnamefont
  {Schlömann}},\ }\bibfield  {title} {\bibinfo {title} {Generation of phonons
  in high‐power ferromagnetic resonance experiments},\ }\href
  {https://doi.org/10.1063/1.1735909} {\bibfield  {journal} {\bibinfo
  {journal} {Journal of Applied Physics}\ }\textbf {\bibinfo {volume} {31}},\
  \bibinfo {pages} {1647} (\bibinfo {year} {1960})}\BibitemShut {NoStop}%
\bibitem [{\citenamefont {Keshtgar}\ \emph {et~al.}(2014)\citenamefont
  {Keshtgar}, \citenamefont {Zareyan},\ and\ \citenamefont
  {Bauer}}]{keshtgar2014acoustic}%
  \BibitemOpen
  \bibfield  {author} {\bibinfo {author} {\bibfnamefont {H.}~\bibnamefont
  {Keshtgar}}, \bibinfo {author} {\bibfnamefont {M.}~\bibnamefont {Zareyan}},\
  and\ \bibinfo {author} {\bibfnamefont {G.~E.}\ \bibnamefont {Bauer}},\
  }\bibfield  {title} {\bibinfo {title} {Acoustic parametric pumping of spin
  waves},\ }\href@noop {} {\bibfield  {journal} {\bibinfo  {journal} {Solid
  state communications}\ }\textbf {\bibinfo {volume} {198}},\ \bibinfo {pages}
  {30} (\bibinfo {year} {2014})}\BibitemShut {NoStop}%
\bibitem [{\citenamefont {Callen}(1968)}]{callen1968magnetostriction}%
  \BibitemOpen
  \bibfield  {author} {\bibinfo {author} {\bibfnamefont {E.}~\bibnamefont
  {Callen}},\ }\bibfield  {title} {\bibinfo {title} {Magnetostriction},\
  }\href@noop {} {\bibfield  {journal} {\bibinfo  {journal} {J. Appl. Phys.}\
  }\textbf {\bibinfo {volume} {39}},\ \bibinfo {pages} {519} (\bibinfo {year}
  {1968})}\BibitemShut {NoStop}%
\bibitem [{\citenamefont {Potts}\ \emph {et~al.}(2020)\citenamefont {Potts},
  \citenamefont {Bittencourt}, \citenamefont {Viola~Kusminskiy},\ and\
  \citenamefont {Davis}}]{potts2020magnon}%
  \BibitemOpen
  \bibfield  {author} {\bibinfo {author} {\bibfnamefont {C.~A.}\ \bibnamefont
  {Potts}}, \bibinfo {author} {\bibfnamefont {V.~A. S.~V.}\ \bibnamefont
  {Bittencourt}}, \bibinfo {author} {\bibfnamefont {S.}~\bibnamefont
  {Viola~Kusminskiy}},\ and\ \bibinfo {author} {\bibfnamefont {J.~P.}\
  \bibnamefont {Davis}},\ }\bibfield  {title} {\bibinfo {title} {Magnon-phonon
  quantum correlation thermometry},\ }\href@noop {} {\bibfield  {journal}
  {\bibinfo  {journal} {Phys. Rev. Appl.}\ }\textbf {\bibinfo {volume} {13}},\
  \bibinfo {pages} {064001} (\bibinfo {year} {2020})}\BibitemShut {NoStop}%
\bibitem [{\citenamefont {Fan}\ \emph {et~al.}(2022)\citenamefont {Fan},
  \citenamefont {Qiu}, \citenamefont {Gr{\"o}blacher},\ and\ \citenamefont
  {Li}}]{fan2022microwave}%
  \BibitemOpen
  \bibfield  {author} {\bibinfo {author} {\bibfnamefont {Z.-Y.}\ \bibnamefont
  {Fan}}, \bibinfo {author} {\bibfnamefont {L.}~\bibnamefont {Qiu}}, \bibinfo
  {author} {\bibfnamefont {S.}~\bibnamefont {Gr{\"o}blacher}},\ and\ \bibinfo
  {author} {\bibfnamefont {J.}~\bibnamefont {Li}},\ }\bibfield  {title}
  {\bibinfo {title} {Microwave-optics entanglement via cavity
  optomagnomechanics},\ }\href@noop {} {\bibfield  {journal} {\bibinfo
  {journal} {arXiv:2208.10703}\ } (\bibinfo {year} {2022})}\BibitemShut
  {NoStop}%
\bibitem [{\citenamefont {Li}\ and\ \citenamefont
  {Gr{\"o}blacher}(2021)}]{li2021entangling}%
  \BibitemOpen
  \bibfield  {author} {\bibinfo {author} {\bibfnamefont {J.}~\bibnamefont
  {Li}}\ and\ \bibinfo {author} {\bibfnamefont {S.}~\bibnamefont
  {Gr{\"o}blacher}},\ }\bibfield  {title} {\bibinfo {title} {Entangling the
  vibrational modes of two massive ferromagnetic spheres using cavity
  magnomechanics},\ }\href@noop {} {\bibfield  {journal} {\bibinfo  {journal}
  {Quantum Sci. Tech.}\ }\textbf {\bibinfo {volume} {6}},\ \bibinfo {pages}
  {024005} (\bibinfo {year} {2021})}\BibitemShut {NoStop}%
\bibitem [{\citenamefont {Li}\ and\ \citenamefont
  {Zhu}(2019)}]{li2019entangling}%
  \BibitemOpen
  \bibfield  {author} {\bibinfo {author} {\bibfnamefont {J.}~\bibnamefont
  {Li}}\ and\ \bibinfo {author} {\bibfnamefont {S.-Y.}\ \bibnamefont {Zhu}},\
  }\bibfield  {title} {\bibinfo {title} {Entangling two magnon modes via
  magnetostrictive interaction},\ }\href@noop {} {\bibfield  {journal}
  {\bibinfo  {journal} {New J. Phys.}\ }\textbf {\bibinfo {volume} {21}},\
  \bibinfo {pages} {085001} (\bibinfo {year} {2019})}\BibitemShut {NoStop}%
\bibitem [{\citenamefont {Li}\ \emph {et~al.}(2018)\citenamefont {Li},
  \citenamefont {Zhu},\ and\ \citenamefont {Agarwal}}]{li2018magnon}%
  \BibitemOpen
  \bibfield  {author} {\bibinfo {author} {\bibfnamefont {J.}~\bibnamefont
  {Li}}, \bibinfo {author} {\bibfnamefont {S.-Y.}\ \bibnamefont {Zhu}},\ and\
  \bibinfo {author} {\bibfnamefont {G.~S.}\ \bibnamefont {Agarwal}},\
  }\bibfield  {title} {\bibinfo {title} {Magnon-photon-phonon entanglement in
  cavity magnomechanics},\ }\href@noop {} {\bibfield  {journal} {\bibinfo
  {journal} {Phys. Rev. Lett.}\ }\textbf {\bibinfo {volume} {121}},\ \bibinfo
  {pages} {203601} (\bibinfo {year} {2018})}\BibitemShut {NoStop}%
\bibitem [{\citenamefont {Li}\ \emph {et~al.}(2021{\natexlab{a}})\citenamefont
  {Li}, \citenamefont {Wang}, \citenamefont {You},\ and\ \citenamefont
  {Zhu}}]{li2021squeezing}%
  \BibitemOpen
  \bibfield  {author} {\bibinfo {author} {\bibfnamefont {J.}~\bibnamefont
  {Li}}, \bibinfo {author} {\bibfnamefont {Y.-P.}\ \bibnamefont {Wang}},
  \bibinfo {author} {\bibfnamefont {J.}~\bibnamefont {You}},\ and\ \bibinfo
  {author} {\bibfnamefont {S.-Y.}\ \bibnamefont {Zhu}},\ }\bibfield  {title}
  {\bibinfo {title} {Squeezing microwave fields via magnetostrictive
  interaction},\ }\href@noop {} {\bibfield  {journal} {\bibinfo  {journal}
  {arXiv:2101.02796}\ } (\bibinfo {year} {2021}{\natexlab{a}})}\BibitemShut
  {NoStop}%
\bibitem [{\citenamefont {Ding}\ \emph {et~al.}(2022)\citenamefont {Ding},
  \citenamefont {Shi}, \citenamefont {Liu},\ and\ \citenamefont
  {Zheng}}]{ding2022magnon}%
  \BibitemOpen
  \bibfield  {author} {\bibinfo {author} {\bibfnamefont {M.-S.}\ \bibnamefont
  {Ding}}, \bibinfo {author} {\bibfnamefont {Y.}~\bibnamefont {Shi}}, \bibinfo
  {author} {\bibfnamefont {Y.-j.}\ \bibnamefont {Liu}},\ and\ \bibinfo {author}
  {\bibfnamefont {L.}~\bibnamefont {Zheng}},\ }\bibfield  {title} {\bibinfo
  {title} {Magnon squeezing enhanced entanglement in a cavity magnomechanical
  system},\ }\href@noop {} {\bibfield  {journal} {\bibinfo  {journal}
  {arXiv:2205.14569}\ } (\bibinfo {year} {2022})}\BibitemShut {NoStop}%
\bibitem [{\citenamefont {Asjad}\ \emph {et~al.}(2022)\citenamefont {Asjad},
  \citenamefont {Li}, \citenamefont {Zhu},\ and\ \citenamefont
  {You}}]{asjad2022magnon}%
  \BibitemOpen
  \bibfield  {author} {\bibinfo {author} {\bibfnamefont {M.}~\bibnamefont
  {Asjad}}, \bibinfo {author} {\bibfnamefont {J.}~\bibnamefont {Li}}, \bibinfo
  {author} {\bibfnamefont {S.-Y.}\ \bibnamefont {Zhu}},\ and\ \bibinfo {author}
  {\bibfnamefont {J.}~\bibnamefont {You}},\ }\bibfield  {title} {\bibinfo
  {title} {Magnon squeezing enhanced ground-state cooling in cavity
  magnomechanics},\ }\href@noop {} {\bibfield  {journal} {\bibinfo  {journal}
  {arXiv:2203.10767}\ } (\bibinfo {year} {2022})}\BibitemShut {NoStop}%
\bibitem [{\citenamefont {Li}\ \emph {et~al.}(2019)\citenamefont {Li},
  \citenamefont {Zhu},\ and\ \citenamefont {Agarwal}}]{li2019squeezed}%
  \BibitemOpen
  \bibfield  {author} {\bibinfo {author} {\bibfnamefont {J.}~\bibnamefont
  {Li}}, \bibinfo {author} {\bibfnamefont {S.-Y.}\ \bibnamefont {Zhu}},\ and\
  \bibinfo {author} {\bibfnamefont {G.~S.}\ \bibnamefont {Agarwal}},\
  }\bibfield  {title} {\bibinfo {title} {Squeezed states of magnons and phonons
  in cavity magnomechanics},\ }\href@noop {} {\bibfield  {journal} {\bibinfo
  {journal} {Phys. Rev. A}\ }\textbf {\bibinfo {volume} {99}},\ \bibinfo
  {pages} {021801(R)} (\bibinfo {year} {2019})}\BibitemShut {NoStop}%
\bibitem [{\citenamefont {Zhang}\ \emph {et~al.}(2021)\citenamefont {Zhang},
  \citenamefont {Wang}, \citenamefont {Bai}, \citenamefont {Wang},
  \citenamefont {Zhang},\ and\ \citenamefont {Wang}}]{zhang2021generation}%
  \BibitemOpen
  \bibfield  {author} {\bibinfo {author} {\bibfnamefont {W.}~\bibnamefont
  {Zhang}}, \bibinfo {author} {\bibfnamefont {D.-Y.}\ \bibnamefont {Wang}},
  \bibinfo {author} {\bibfnamefont {C.-H.}\ \bibnamefont {Bai}}, \bibinfo
  {author} {\bibfnamefont {T.}~\bibnamefont {Wang}}, \bibinfo {author}
  {\bibfnamefont {S.}~\bibnamefont {Zhang}},\ and\ \bibinfo {author}
  {\bibfnamefont {H.-F.}\ \bibnamefont {Wang}},\ }\bibfield  {title} {\bibinfo
  {title} {Generation and transfer of squeezed states in a cavity
  magnomechanical system by two-tone microwave fields},\ }\href@noop {}
  {\bibfield  {journal} {\bibinfo  {journal} {Opt. Express}\ }\textbf {\bibinfo
  {volume} {29}},\ \bibinfo {pages} {11773} (\bibinfo {year}
  {2021})}\BibitemShut {NoStop}%
\bibitem [{\citenamefont {Sarma}\ \emph {et~al.}(2021)\citenamefont {Sarma},
  \citenamefont {Busch},\ and\ \citenamefont {Twamley}}]{sarma2021cavity}%
  \BibitemOpen
  \bibfield  {author} {\bibinfo {author} {\bibfnamefont {B.}~\bibnamefont
  {Sarma}}, \bibinfo {author} {\bibfnamefont {T.}~\bibnamefont {Busch}},\ and\
  \bibinfo {author} {\bibfnamefont {J.}~\bibnamefont {Twamley}},\ }\bibfield
  {title} {\bibinfo {title} {Cavity magnomechanical storage and retrieval of
  quantum states},\ }\href@noop {} {\bibfield  {journal} {\bibinfo  {journal}
  {New J. Phys.}\ }\textbf {\bibinfo {volume} {23}},\ \bibinfo {pages} {043041}
  (\bibinfo {year} {2021})}\BibitemShut {NoStop}%
\bibitem [{\citenamefont {Li}\ \emph {et~al.}(2020)\citenamefont {Li},
  \citenamefont {Yang}, \citenamefont {Shui}, \citenamefont {Li}, \citenamefont
  {Wang},\ and\ \citenamefont {Wu}}]{li2020phase}%
  \BibitemOpen
  \bibfield  {author} {\bibinfo {author} {\bibfnamefont {X.}~\bibnamefont
  {Li}}, \bibinfo {author} {\bibfnamefont {W.-X.}\ \bibnamefont {Yang}},
  \bibinfo {author} {\bibfnamefont {T.}~\bibnamefont {Shui}}, \bibinfo {author}
  {\bibfnamefont {L.}~\bibnamefont {Li}}, \bibinfo {author} {\bibfnamefont
  {X.}~\bibnamefont {Wang}},\ and\ \bibinfo {author} {\bibfnamefont
  {Z.}~\bibnamefont {Wu}},\ }\bibfield  {title} {\bibinfo {title} {Phase
  control of the transmission in cavity magnomechanical system with magnon
  driving},\ }\href@noop {} {\bibfield  {journal} {\bibinfo  {journal} {J.
  Appl. Phys.}\ }\textbf {\bibinfo {volume} {128}},\ \bibinfo {pages} {233101}
  (\bibinfo {year} {2020})}\BibitemShut {NoStop}%
\bibitem [{\citenamefont {Kong}\ \emph {et~al.}(2019)\citenamefont {Kong},
  \citenamefont {Wang}, \citenamefont {Liu}, \citenamefont {Xiong},\ and\
  \citenamefont {Wu}}]{kong2019magnetically}%
  \BibitemOpen
  \bibfield  {author} {\bibinfo {author} {\bibfnamefont {C.}~\bibnamefont
  {Kong}}, \bibinfo {author} {\bibfnamefont {B.}~\bibnamefont {Wang}}, \bibinfo
  {author} {\bibfnamefont {Z.-X.}\ \bibnamefont {Liu}}, \bibinfo {author}
  {\bibfnamefont {H.}~\bibnamefont {Xiong}},\ and\ \bibinfo {author}
  {\bibfnamefont {Y.}~\bibnamefont {Wu}},\ }\bibfield  {title} {\bibinfo
  {title} {Magnetically controllable slow light based on magnetostrictive
  forces},\ }\href@noop {} {\bibfield  {journal} {\bibinfo  {journal} {Opt.
  Express}\ }\textbf {\bibinfo {volume} {27}},\ \bibinfo {pages} {5544}
  (\bibinfo {year} {2019})}\BibitemShut {NoStop}%
\bibitem [{\citenamefont {Li}\ \emph {et~al.}(2021{\natexlab{b}})\citenamefont
  {Li}, \citenamefont {Wang}, \citenamefont {Wu}, \citenamefont {Zhu},\ and\
  \citenamefont {You}}]{li2021quantum}%
  \BibitemOpen
  \bibfield  {author} {\bibinfo {author} {\bibfnamefont {J.}~\bibnamefont
  {Li}}, \bibinfo {author} {\bibfnamefont {Y.-P.}\ \bibnamefont {Wang}},
  \bibinfo {author} {\bibfnamefont {W.-J.}\ \bibnamefont {Wu}}, \bibinfo
  {author} {\bibfnamefont {S.-Y.}\ \bibnamefont {Zhu}},\ and\ \bibinfo {author}
  {\bibfnamefont {J.~Q.}\ \bibnamefont {You}},\ }\bibfield  {title} {\bibinfo
  {title} {Quantum network with magnonic and mechanical nodes},\ }\href@noop {}
  {\bibfield  {journal} {\bibinfo  {journal} {PRX Quantum}\ }\textbf {\bibinfo
  {volume} {2}},\ \bibinfo {pages} {040344} (\bibinfo {year}
  {2021}{\natexlab{b}})}\BibitemShut {NoStop}%
\bibitem [{\citenamefont {Hatanaka}\ \emph {et~al.}(2022)\citenamefont
  {Hatanaka}, \citenamefont {Asano}, \citenamefont {Okamoto}, \citenamefont
  {Kunihashi}, \citenamefont {Sanada},\ and\ \citenamefont
  {Yamaguchi}}]{hatanaka2022chip}%
  \BibitemOpen
  \bibfield  {author} {\bibinfo {author} {\bibfnamefont {D.}~\bibnamefont
  {Hatanaka}}, \bibinfo {author} {\bibfnamefont {M.}~\bibnamefont {Asano}},
  \bibinfo {author} {\bibfnamefont {H.}~\bibnamefont {Okamoto}}, \bibinfo
  {author} {\bibfnamefont {Y.}~\bibnamefont {Kunihashi}}, \bibinfo {author}
  {\bibfnamefont {H.}~\bibnamefont {Sanada}},\ and\ \bibinfo {author}
  {\bibfnamefont {H.}~\bibnamefont {Yamaguchi}},\ }\bibfield  {title} {\bibinfo
  {title} {On-chip coherent transduction between magnons and acoustic phonons
  in cavity magnomechanics},\ }\href@noop {} {\bibfield  {journal} {\bibinfo
  {journal} {Phys. Rev. Appl.}\ }\textbf {\bibinfo {volume} {17}},\ \bibinfo
  {pages} {034024} (\bibinfo {year} {2022})}\BibitemShut {NoStop}%
\bibitem [{\citenamefont {Clark}\ \emph {et~al.}(2017)\citenamefont {Clark},
  \citenamefont {Lecocq}, \citenamefont {Simmonds}, \citenamefont {Aumentado},\
  and\ \citenamefont {Teufel}}]{clark2017sideband}%
  \BibitemOpen
  \bibfield  {author} {\bibinfo {author} {\bibfnamefont {J.~B.}\ \bibnamefont
  {Clark}}, \bibinfo {author} {\bibfnamefont {F.}~\bibnamefont {Lecocq}},
  \bibinfo {author} {\bibfnamefont {R.~W.}\ \bibnamefont {Simmonds}}, \bibinfo
  {author} {\bibfnamefont {J.}~\bibnamefont {Aumentado}},\ and\ \bibinfo
  {author} {\bibfnamefont {J.~D.}\ \bibnamefont {Teufel}},\ }\bibfield  {title}
  {\bibinfo {title} {Sideband cooling beyond the quantum backaction limit with
  squeezed light},\ }\href@noop {} {\bibfield  {journal} {\bibinfo  {journal}
  {Nature}\ }\textbf {\bibinfo {volume} {541}},\ \bibinfo {pages} {191}
  (\bibinfo {year} {2017})}\BibitemShut {NoStop}%
\bibitem [{\citenamefont {Purdy}\ \emph {et~al.}(2017)\citenamefont {Purdy},
  \citenamefont {Grutter}, \citenamefont {Srinivasan},\ and\ \citenamefont
  {Taylor}}]{purdy2017quantum}%
  \BibitemOpen
  \bibfield  {author} {\bibinfo {author} {\bibfnamefont {T.}~\bibnamefont
  {Purdy}}, \bibinfo {author} {\bibfnamefont {K.}~\bibnamefont {Grutter}},
  \bibinfo {author} {\bibfnamefont {K.}~\bibnamefont {Srinivasan}},\ and\
  \bibinfo {author} {\bibfnamefont {J.}~\bibnamefont {Taylor}},\ }\bibfield
  {title} {\bibinfo {title} {Quantum correlations from a room-temperature
  optomechanical cavity},\ }\href@noop {} {\bibfield  {journal} {\bibinfo
  {journal} {Science}\ }\textbf {\bibinfo {volume} {356}},\ \bibinfo {pages}
  {1265} (\bibinfo {year} {2017})}\BibitemShut {NoStop}%
\bibitem [{\citenamefont {Hertzberg}\ \emph {et~al.}(2010)\citenamefont
  {Hertzberg}, \citenamefont {Rocheleau}, \citenamefont {Ndukum}, \citenamefont
  {Savva}, \citenamefont {Clerk},\ and\ \citenamefont
  {Schwab}}]{hertzberg2010back}%
  \BibitemOpen
  \bibfield  {author} {\bibinfo {author} {\bibfnamefont {J.}~\bibnamefont
  {Hertzberg}}, \bibinfo {author} {\bibfnamefont {T.}~\bibnamefont
  {Rocheleau}}, \bibinfo {author} {\bibfnamefont {T.}~\bibnamefont {Ndukum}},
  \bibinfo {author} {\bibfnamefont {M.}~\bibnamefont {Savva}}, \bibinfo
  {author} {\bibfnamefont {A.~A.}\ \bibnamefont {Clerk}},\ and\ \bibinfo
  {author} {\bibfnamefont {K.}~\bibnamefont {Schwab}},\ }\bibfield  {title}
  {\bibinfo {title} {Back-action-evading measurements of nanomechanical
  motion},\ }\href@noop {} {\bibfield  {journal} {\bibinfo  {journal} {Nat.
  Phys.}\ }\textbf {\bibinfo {volume} {6}},\ \bibinfo {pages} {213} (\bibinfo
  {year} {2010})}\BibitemShut {NoStop}%
\bibitem [{\citenamefont {Teufel}\ \emph {et~al.}(2016)\citenamefont {Teufel},
  \citenamefont {Lecocq},\ and\ \citenamefont
  {Simmonds}}]{teufel2016overwhelming}%
  \BibitemOpen
  \bibfield  {author} {\bibinfo {author} {\bibfnamefont {J.~D.}\ \bibnamefont
  {Teufel}}, \bibinfo {author} {\bibfnamefont {F.}~\bibnamefont {Lecocq}},\
  and\ \bibinfo {author} {\bibfnamefont {R.~W.}\ \bibnamefont {Simmonds}},\
  }\bibfield  {title} {\bibinfo {title} {Overwhelming thermomechanical motion
  with microwave radiation pressure shot noise},\ }\href
  {https://doi.org/10.1103/PhysRevLett.116.013602} {\bibfield  {journal}
  {\bibinfo  {journal} {Phys. Rev. Lett.}\ }\textbf {\bibinfo {volume} {116}},\
  \bibinfo {pages} {013602} (\bibinfo {year} {2016})}\BibitemShut {NoStop}%
\bibitem [{\citenamefont {Cripe}\ \emph {et~al.}(2019)\citenamefont {Cripe},
  \citenamefont {Aggarwal}, \citenamefont {Lanza}, \citenamefont {Libson},
  \citenamefont {Singh}, \citenamefont {Heu}, \citenamefont {Follman},
  \citenamefont {Cole}, \citenamefont {Mavalvala},\ and\ \citenamefont
  {Corbitt}}]{cripe2019measurement}%
  \BibitemOpen
  \bibfield  {author} {\bibinfo {author} {\bibfnamefont {J.}~\bibnamefont
  {Cripe}}, \bibinfo {author} {\bibfnamefont {N.}~\bibnamefont {Aggarwal}},
  \bibinfo {author} {\bibfnamefont {R.}~\bibnamefont {Lanza}}, \bibinfo
  {author} {\bibfnamefont {A.}~\bibnamefont {Libson}}, \bibinfo {author}
  {\bibfnamefont {R.}~\bibnamefont {Singh}}, \bibinfo {author} {\bibfnamefont
  {P.}~\bibnamefont {Heu}}, \bibinfo {author} {\bibfnamefont {D.}~\bibnamefont
  {Follman}}, \bibinfo {author} {\bibfnamefont {G.~D.}\ \bibnamefont {Cole}},
  \bibinfo {author} {\bibfnamefont {N.}~\bibnamefont {Mavalvala}},\ and\
  \bibinfo {author} {\bibfnamefont {T.}~\bibnamefont {Corbitt}},\ }\bibfield
  {title} {\bibinfo {title} {Measurement of quantum back action in the audio
  band at room temperature},\ }\href@noop {} {\bibfield  {journal} {\bibinfo
  {journal} {Nature}\ }\textbf {\bibinfo {volume} {568}},\ \bibinfo {pages}
  {364} (\bibinfo {year} {2019})}\BibitemShut {NoStop}%
\bibitem [{\citenamefont {Purdy}\ \emph {et~al.}(2013)\citenamefont {Purdy},
  \citenamefont {Peterson},\ and\ \citenamefont
  {Regal}}]{purdy2013observation}%
  \BibitemOpen
  \bibfield  {author} {\bibinfo {author} {\bibfnamefont {T.~P.}\ \bibnamefont
  {Purdy}}, \bibinfo {author} {\bibfnamefont {R.~W.}\ \bibnamefont
  {Peterson}},\ and\ \bibinfo {author} {\bibfnamefont {C.}~\bibnamefont
  {Regal}},\ }\bibfield  {title} {\bibinfo {title} {Observation of radiation
  pressure shot noise on a macroscopic object},\ }\href@noop {} {\bibfield
  {journal} {\bibinfo  {journal} {Science}\ }\textbf {\bibinfo {volume}
  {339}},\ \bibinfo {pages} {801} (\bibinfo {year} {2013})}\BibitemShut
  {NoStop}%
\bibitem [{\citenamefont {Gloppe}\ \emph {et~al.}(2019)\citenamefont {Gloppe},
  \citenamefont {Hisatomi}, \citenamefont {Nakata}, \citenamefont {Nakamura},\
  and\ \citenamefont {Usami}}]{gloppe2019resonant}%
  \BibitemOpen
  \bibfield  {author} {\bibinfo {author} {\bibfnamefont {A.}~\bibnamefont
  {Gloppe}}, \bibinfo {author} {\bibfnamefont {R.}~\bibnamefont {Hisatomi}},
  \bibinfo {author} {\bibfnamefont {Y.}~\bibnamefont {Nakata}}, \bibinfo
  {author} {\bibfnamefont {Y.}~\bibnamefont {Nakamura}},\ and\ \bibinfo
  {author} {\bibfnamefont {K.}~\bibnamefont {Usami}},\ }\bibfield  {title}
  {\bibinfo {title} {Resonant magnetic induction tomography of a magnetized
  sphere},\ }\href@noop {} {\bibfield  {journal} {\bibinfo  {journal} {Phys.
  Rev. Appl.}\ }\textbf {\bibinfo {volume} {12}},\ \bibinfo {pages} {014061}
  (\bibinfo {year} {2019})}\BibitemShut {NoStop}%
\bibitem [{\citenamefont {Bittencourt}\ \emph {et~al.}(2023)\citenamefont
  {Bittencourt}, \citenamefont {Potts}, \citenamefont {Huang}, \citenamefont
  {Davis},\ and\ \citenamefont {Viola~Kusminskiy}}]{BittencourtDamping2022}%
  \BibitemOpen
  \bibfield  {author} {\bibinfo {author} {\bibfnamefont {V.~A. S.~V.}\
  \bibnamefont {Bittencourt}}, \bibinfo {author} {\bibfnamefont {C.~A.}\
  \bibnamefont {Potts}}, \bibinfo {author} {\bibfnamefont {Y.}~\bibnamefont
  {Huang}}, \bibinfo {author} {\bibfnamefont {J.~P.}\ \bibnamefont {Davis}},\
  and\ \bibinfo {author} {\bibfnamefont {S.}~\bibnamefont {Viola~Kusminskiy}},\
  }\bibfield  {title} {\bibinfo {title} {Magnomechanical backaction corrections
  due to coupling to higher order walker modes and kerr nonlinearities},\
  }\href@noop {} {\bibfield  {journal} {\bibinfo  {journal} {arXiv:2301.11920}\
  } (\bibinfo {year} {2023})}\BibitemShut {NoStop}%
\bibitem [{\citenamefont {Braginsky}\ \emph {et~al.}(1980)\citenamefont
  {Braginsky}, \citenamefont {Vorontsov},\ and\ \citenamefont
  {Thorne}}]{braginsky1980quantum}%
  \BibitemOpen
  \bibfield  {author} {\bibinfo {author} {\bibfnamefont {V.~B.}\ \bibnamefont
  {Braginsky}}, \bibinfo {author} {\bibfnamefont {Y.~I.}\ \bibnamefont
  {Vorontsov}},\ and\ \bibinfo {author} {\bibfnamefont {K.~S.}\ \bibnamefont
  {Thorne}},\ }\bibfield  {title} {\bibinfo {title} {Quantum nondemolition
  measurements},\ }\href@noop {} {\bibfield  {journal} {\bibinfo  {journal}
  {Science}\ }\textbf {\bibinfo {volume} {209}},\ \bibinfo {pages} {547}
  (\bibinfo {year} {1980})}\BibitemShut {NoStop}%
\bibitem [{\citenamefont {Braginsky}\ and\ \citenamefont
  {Khalili}(1995)}]{braginsky1995quantum}%
  \BibitemOpen
  \bibfield  {author} {\bibinfo {author} {\bibfnamefont {V.~B.}\ \bibnamefont
  {Braginsky}}\ and\ \bibinfo {author} {\bibfnamefont {F.~Y.}\ \bibnamefont
  {Khalili}},\ }\href@noop {} {\emph {\bibinfo {title} {Quantum measurement}}}\
  (\bibinfo  {publisher} {Cambridge University Press},\ \bibinfo {year}
  {1995})\BibitemShut {NoStop}%
\bibitem [{Sup()}]{SupplimentaryInfo}%
  \BibitemOpen
  \bibfield  {title} {\bibinfo {title} {See supplemental material at},\
  }\href@noop {} {\ }\BibitemShut {NoStop}%
\bibitem [{\citenamefont {Lecocq}\ \emph {et~al.}(2015)\citenamefont {Lecocq},
  \citenamefont {Clark}, \citenamefont {Simmonds}, \citenamefont {Aumentado},\
  and\ \citenamefont {Teufel}}]{lecocq2015quantum}%
  \BibitemOpen
  \bibfield  {author} {\bibinfo {author} {\bibfnamefont {F.}~\bibnamefont
  {Lecocq}}, \bibinfo {author} {\bibfnamefont {J.~B.}\ \bibnamefont {Clark}},
  \bibinfo {author} {\bibfnamefont {R.~W.}\ \bibnamefont {Simmonds}}, \bibinfo
  {author} {\bibfnamefont {J.}~\bibnamefont {Aumentado}},\ and\ \bibinfo
  {author} {\bibfnamefont {J.~D.}\ \bibnamefont {Teufel}},\ }\bibfield  {title}
  {\bibinfo {title} {Quantum nondemolition measurement of a nonclassical state
  of a massive object},\ }\href@noop {} {\bibfield  {journal} {\bibinfo
  {journal} {Phys. Rev. X}\ }\textbf {\bibinfo {volume} {5}},\ \bibinfo {pages}
  {041037} (\bibinfo {year} {2015})}\BibitemShut {NoStop}%
\bibitem [{\citenamefont {Clerk}\ \emph {et~al.}(2008)\citenamefont {Clerk},
  \citenamefont {Marquardt},\ and\ \citenamefont {Jacobs}}]{clerk2008back}%
  \BibitemOpen
  \bibfield  {author} {\bibinfo {author} {\bibfnamefont {A.~A.}\ \bibnamefont
  {Clerk}}, \bibinfo {author} {\bibfnamefont {F.}~\bibnamefont {Marquardt}},\
  and\ \bibinfo {author} {\bibfnamefont {K.}~\bibnamefont {Jacobs}},\
  }\bibfield  {title} {\bibinfo {title} {Back-action evasion and squeezing of a
  mechanical resonator using a cavity detector},\ }\href@noop {} {\bibfield
  {journal} {\bibinfo  {journal} {New J. Phys.}\ }\textbf {\bibinfo {volume}
  {10}},\ \bibinfo {pages} {095010} (\bibinfo {year} {2008})}\BibitemShut
  {NoStop}%
\bibitem [{\citenamefont {Suh}\ \emph {et~al.}(2014)\citenamefont {Suh},
  \citenamefont {Weinstein}, \citenamefont {Lei}, \citenamefont {Wollman},
  \citenamefont {Steinke}, \citenamefont {Meystre}, \citenamefont {Clerk},\
  and\ \citenamefont {Schwab}}]{suh2014mechanically}%
  \BibitemOpen
  \bibfield  {author} {\bibinfo {author} {\bibfnamefont {J.}~\bibnamefont
  {Suh}}, \bibinfo {author} {\bibfnamefont {A.}~\bibnamefont {Weinstein}},
  \bibinfo {author} {\bibfnamefont {C.}~\bibnamefont {Lei}}, \bibinfo {author}
  {\bibfnamefont {E.}~\bibnamefont {Wollman}}, \bibinfo {author} {\bibfnamefont
  {S.}~\bibnamefont {Steinke}}, \bibinfo {author} {\bibfnamefont
  {P.}~\bibnamefont {Meystre}}, \bibinfo {author} {\bibfnamefont {A.~A.}\
  \bibnamefont {Clerk}},\ and\ \bibinfo {author} {\bibfnamefont
  {K.}~\bibnamefont {Schwab}},\ }\bibfield  {title} {\bibinfo {title}
  {Mechanically detecting and avoiding the quantum fluctuations of a microwave
  field},\ }\href@noop {} {\bibfield  {journal} {\bibinfo  {journal} {Science}\
  }\textbf {\bibinfo {volume} {344}},\ \bibinfo {pages} {1262} (\bibinfo {year}
  {2014})}\BibitemShut {NoStop}%
\bibitem [{\citenamefont {Shomroni}\ \emph {et~al.}(2019)\citenamefont
  {Shomroni}, \citenamefont {Qiu}, \citenamefont {Malz}, \citenamefont
  {Nunnenkamp},\ and\ \citenamefont {Kippenberg}}]{shomroni2019optical}%
  \BibitemOpen
  \bibfield  {author} {\bibinfo {author} {\bibfnamefont {I.}~\bibnamefont
  {Shomroni}}, \bibinfo {author} {\bibfnamefont {L.}~\bibnamefont {Qiu}},
  \bibinfo {author} {\bibfnamefont {D.}~\bibnamefont {Malz}}, \bibinfo {author}
  {\bibfnamefont {A.}~\bibnamefont {Nunnenkamp}},\ and\ \bibinfo {author}
  {\bibfnamefont {T.~J.}\ \bibnamefont {Kippenberg}},\ }\bibfield  {title}
  {\bibinfo {title} {Optical backaction-evading measurement of a mechanical
  oscillator},\ }\href@noop {} {\bibfield  {journal} {\bibinfo  {journal} {Nat.
  Commun.}\ }\textbf {\bibinfo {volume} {10}},\ \bibinfo {pages} {1} (\bibinfo
  {year} {2019})}\BibitemShut {NoStop}%
\bibitem [{\citenamefont {Ockeloen-Korppi}\ \emph {et~al.}(2016)\citenamefont
  {Ockeloen-Korppi}, \citenamefont {Damsk{\"a}gg}, \citenamefont
  {Pirkkalainen}, \citenamefont {Clerk}, \citenamefont {Woolley},\ and\
  \citenamefont {Sillanp{\"a}{\"a}}}]{ockeloen2016quantum}%
  \BibitemOpen
  \bibfield  {author} {\bibinfo {author} {\bibfnamefont {C.~F.}\ \bibnamefont
  {Ockeloen-Korppi}}, \bibinfo {author} {\bibfnamefont {E.}~\bibnamefont
  {Damsk{\"a}gg}}, \bibinfo {author} {\bibfnamefont {J.-M.}\ \bibnamefont
  {Pirkkalainen}}, \bibinfo {author} {\bibfnamefont {A.~A.}\ \bibnamefont
  {Clerk}}, \bibinfo {author} {\bibfnamefont {M.~J.}\ \bibnamefont {Woolley}},\
  and\ \bibinfo {author} {\bibfnamefont {M.~A.}\ \bibnamefont
  {Sillanp{\"a}{\"a}}},\ }\bibfield  {title} {\bibinfo {title} {Quantum
  backaction evading measurement of collective mechanical modes},\ }\href@noop
  {} {\bibfield  {journal} {\bibinfo  {journal} {Phys. Rev. Lett.}\ }\textbf
  {\bibinfo {volume} {117}},\ \bibinfo {pages} {140401} (\bibinfo {year}
  {2016})}\BibitemShut {NoStop}%
\bibitem [{\citenamefont {Wang}\ \emph {et~al.}(2018)\citenamefont {Wang},
  \citenamefont {Zhang}, \citenamefont {Zhang}, \citenamefont {Li},
  \citenamefont {Hu},\ and\ \citenamefont {You}}]{wang2018bistability}%
  \BibitemOpen
  \bibfield  {author} {\bibinfo {author} {\bibfnamefont {Y.-P.}\ \bibnamefont
  {Wang}}, \bibinfo {author} {\bibfnamefont {G.-Q.}\ \bibnamefont {Zhang}},
  \bibinfo {author} {\bibfnamefont {D.}~\bibnamefont {Zhang}}, \bibinfo
  {author} {\bibfnamefont {T.-F.}\ \bibnamefont {Li}}, \bibinfo {author}
  {\bibfnamefont {C.-M.}\ \bibnamefont {Hu}},\ and\ \bibinfo {author}
  {\bibfnamefont {J.~Q.}\ \bibnamefont {You}},\ }\bibfield  {title} {\bibinfo
  {title} {Bistability of cavity magnon polaritons},\ }\href@noop {} {\bibfield
   {journal} {\bibinfo  {journal} {Phys. Rev. Lett.}\ }\textbf {\bibinfo
  {volume} {120}},\ \bibinfo {pages} {057202} (\bibinfo {year}
  {2018})}\BibitemShut {NoStop}%
\bibitem [{\citenamefont {Lake}\ \emph {et~al.}(2020)\citenamefont {Lake},
  \citenamefont {Mitchell}, \citenamefont {Sanders},\ and\ \citenamefont
  {Barclay}}]{lake2020two}%
  \BibitemOpen
  \bibfield  {author} {\bibinfo {author} {\bibfnamefont {D.~P.}\ \bibnamefont
  {Lake}}, \bibinfo {author} {\bibfnamefont {M.}~\bibnamefont {Mitchell}},
  \bibinfo {author} {\bibfnamefont {B.~C.}\ \bibnamefont {Sanders}},\ and\
  \bibinfo {author} {\bibfnamefont {P.~E.}\ \bibnamefont {Barclay}},\
  }\bibfield  {title} {\bibinfo {title} {Two-colour interferometry and
  switching through optomechanical dark mode excitation},\ }\href@noop {}
  {\bibfield  {journal} {\bibinfo  {journal} {Nat. Commun.}\ }\textbf {\bibinfo
  {volume} {11}},\ \bibinfo {pages} {1} (\bibinfo {year} {2020})}\BibitemShut
  {NoStop}%
\bibitem [{\citenamefont {LeCraw}\ \emph {et~al.}(1958)\citenamefont {LeCraw},
  \citenamefont {Spencer},\ and\ \citenamefont {Porter}}]{LeCraw1958nonlinear}%
  \BibitemOpen
  \bibfield  {author} {\bibinfo {author} {\bibfnamefont {R.~C.}\ \bibnamefont
  {LeCraw}}, \bibinfo {author} {\bibfnamefont {E.~G.}\ \bibnamefont
  {Spencer}},\ and\ \bibinfo {author} {\bibfnamefont {C.~S.}\ \bibnamefont
  {Porter}},\ }\bibfield  {title} {\bibinfo {title} {Ferromagnetic resonance
  and nonlinear effects in yttrium iron garnet},\ }\href
  {https://doi.org/10.1063/1.1723119} {\bibfield  {journal} {\bibinfo
  {journal} {Journal of Applied Physics}\ }\textbf {\bibinfo {volume} {29}},\
  \bibinfo {pages} {326} (\bibinfo {year} {1958})}\BibitemShut {NoStop}%
\bibitem [{\citenamefont {Kotler}\ \emph {et~al.}(2021)\citenamefont {Kotler},
  \citenamefont {Peterson}, \citenamefont {Shojaee}, \citenamefont {Lecocq},
  \citenamefont {Cicak}, \citenamefont {Kwiatkowski}, \citenamefont {Geller},
  \citenamefont {Glancy}, \citenamefont {Knill}, \citenamefont {Simmonds},
  \citenamefont {Aumentado},\ and\ \citenamefont {Teufel}}]{kotler2021direct}%
  \BibitemOpen
  \bibfield  {author} {\bibinfo {author} {\bibfnamefont {S.}~\bibnamefont
  {Kotler}}, \bibinfo {author} {\bibfnamefont {G.~A.}\ \bibnamefont
  {Peterson}}, \bibinfo {author} {\bibfnamefont {E.}~\bibnamefont {Shojaee}},
  \bibinfo {author} {\bibfnamefont {F.}~\bibnamefont {Lecocq}}, \bibinfo
  {author} {\bibfnamefont {K.}~\bibnamefont {Cicak}}, \bibinfo {author}
  {\bibfnamefont {A.}~\bibnamefont {Kwiatkowski}}, \bibinfo {author}
  {\bibfnamefont {S.}~\bibnamefont {Geller}}, \bibinfo {author} {\bibfnamefont
  {S.}~\bibnamefont {Glancy}}, \bibinfo {author} {\bibfnamefont
  {E.}~\bibnamefont {Knill}}, \bibinfo {author} {\bibfnamefont {R.~W.}\
  \bibnamefont {Simmonds}}, \bibinfo {author} {\bibfnamefont {J.}~\bibnamefont
  {Aumentado}},\ and\ \bibinfo {author} {\bibfnamefont {J.~D.}\ \bibnamefont
  {Teufel}},\ }\bibfield  {title} {\bibinfo {title} {Direct observation of
  deterministic macroscopic entanglement},\ }\href@noop {} {\bibfield
  {journal} {\bibinfo  {journal} {Science}\ }\textbf {\bibinfo {volume}
  {372}},\ \bibinfo {pages} {622} (\bibinfo {year} {2021})}\BibitemShut
  {NoStop}%
\bibitem [{\citenamefont {Rodrigues}\ \emph {et~al.}(2019)\citenamefont
  {Rodrigues}, \citenamefont {Bothner},\ and\ \citenamefont
  {Steele}}]{rodrigues2019coupling}%
  \BibitemOpen
  \bibfield  {author} {\bibinfo {author} {\bibfnamefont {I.}~\bibnamefont
  {Rodrigues}}, \bibinfo {author} {\bibfnamefont {D.}~\bibnamefont {Bothner}},\
  and\ \bibinfo {author} {\bibfnamefont {G.}~\bibnamefont {Steele}},\
  }\bibfield  {title} {\bibinfo {title} {Coupling microwave photons to a
  mechanical resonator using quantum interference},\ }\href@noop {} {\bibfield
  {journal} {\bibinfo  {journal} {Nat. Commun.}\ }\textbf {\bibinfo {volume}
  {10}},\ \bibinfo {pages} {1} (\bibinfo {year} {2019})}\BibitemShut {NoStop}%
\bibitem [{\citenamefont {Penrose}(1996)}]{penrose1996gravity}%
  \BibitemOpen
  \bibfield  {author} {\bibinfo {author} {\bibfnamefont {R.}~\bibnamefont
  {Penrose}},\ }\bibfield  {title} {\bibinfo {title} {On gravity's role in
  quantum state reduction},\ }\href@noop {} {\bibfield  {journal} {\bibinfo
  {journal} {Gen. Relativ. Gravit.}\ }\textbf {\bibinfo {volume} {28}},\
  \bibinfo {pages} {581} (\bibinfo {year} {1996})}\BibitemShut {NoStop}%
\bibitem [{\citenamefont {Diosi}(1987)}]{diosi1987universal}%
  \BibitemOpen
  \bibfield  {author} {\bibinfo {author} {\bibfnamefont {L.}~\bibnamefont
  {Diosi}},\ }\bibfield  {title} {\bibinfo {title} {A universal master equation
  for the gravitational violation of quantum mechanics},\ }\href@noop {}
  {\bibfield  {journal} {\bibinfo  {journal} {Phys. Rev. A}\ }\textbf {\bibinfo
  {volume} {120}},\ \bibinfo {pages} {377} (\bibinfo {year}
  {1987})}\BibitemShut {NoStop}%
\bibitem [{\citenamefont {Kanari-Naish}\ \emph {et~al.}(2021)\citenamefont
  {Kanari-Naish}, \citenamefont {Clarke}, \citenamefont {Vanner},\ and\
  \citenamefont {Laird}}]{kanari2021can}%
  \BibitemOpen
  \bibfield  {author} {\bibinfo {author} {\bibfnamefont {L.~A.}\ \bibnamefont
  {Kanari-Naish}}, \bibinfo {author} {\bibfnamefont {J.}~\bibnamefont
  {Clarke}}, \bibinfo {author} {\bibfnamefont {M.~R.}\ \bibnamefont {Vanner}},\
  and\ \bibinfo {author} {\bibfnamefont {E.~A.}\ \bibnamefont {Laird}},\
  }\bibfield  {title} {\bibinfo {title} {Can the displacemon device test
  objective collapse models?},\ }\href@noop {} {\bibfield  {journal} {\bibinfo
  {journal} {AVS Quantum Sci.}\ }\textbf {\bibinfo {volume} {3}},\ \bibinfo
  {pages} {045603} (\bibinfo {year} {2021})}\BibitemShut {NoStop}%
\bibitem [{\citenamefont {Gely}\ and\ \citenamefont
  {Steele}(2021)}]{gely2021superconducting}%
  \BibitemOpen
  \bibfield  {author} {\bibinfo {author} {\bibfnamefont {M.~F.}\ \bibnamefont
  {Gely}}\ and\ \bibinfo {author} {\bibfnamefont {G.~A.}\ \bibnamefont
  {Steele}},\ }\bibfield  {title} {\bibinfo {title} {Superconducting
  electro-mechanics to test di{\'o}si--penrose effects of general relativity in
  massive superpositions},\ }\href@noop {} {\bibfield  {journal} {\bibinfo
  {journal} {AVS Quantum Sci.}\ }\textbf {\bibinfo {volume} {3}},\ \bibinfo
  {pages} {035601} (\bibinfo {year} {2021})}\BibitemShut {NoStop}%
\bibitem [{\citenamefont {Bild}\ \emph {et~al.}(2022)\citenamefont {Bild},
  \citenamefont {Fadel}, \citenamefont {Yang}, \citenamefont {von L{\"u}pke},
  \citenamefont {Martin}, \citenamefont {Bruno},\ and\ \citenamefont
  {Chu}}]{Marius_Schrodinger_2022}%
  \BibitemOpen
  \bibfield  {author} {\bibinfo {author} {\bibfnamefont {M.}~\bibnamefont
  {Bild}}, \bibinfo {author} {\bibfnamefont {M.}~\bibnamefont {Fadel}},
  \bibinfo {author} {\bibfnamefont {Y.}~\bibnamefont {Yang}}, \bibinfo {author}
  {\bibfnamefont {U.}~\bibnamefont {von L{\"u}pke}}, \bibinfo {author}
  {\bibfnamefont {P.}~\bibnamefont {Martin}}, \bibinfo {author} {\bibfnamefont
  {A.}~\bibnamefont {Bruno}},\ and\ \bibinfo {author} {\bibfnamefont
  {Y.}~\bibnamefont {Chu}},\ }\bibfield  {title} {\bibinfo {title} {Schrodinger
  cat states of a 16-microgram mechanical oscillator},\ }\href@noop {}
  {\bibfield  {journal} {\bibinfo  {journal} {arXiv:2211.00449}\ } (\bibinfo
  {year} {2022})}\BibitemShut {NoStop}%
\end{thebibliography}%

%%%%%%%%%% Merge with supplemental materials %%%%%%%%%%
\pagebreak
\widetext
\begin{center}
\textbf{\large Dynamical Backaction Evading Magnomechanics --- Supplemental Materials}
%\textbf{\large Triple-Resonance Backaction Evasion --- Supplemental Materials}

\vspace{11pt}

C.A. Potts,$^{1,*}$ Y. Huang,$^1$ V.A.S.V. Bittencourt,$^2$ S. {Viola Kusminskiy},$^{3,4}$ J.P. Davis$^{1,\dagger}$

\vspace{11pt}

\footnotesize
$^1$ \textit{Department of Physics, University of Alberta, Edmonton, Alberta T6G 2E9, Canada}

$^2$ \textit{Max Planck Institute for the Science of Light, Staudtstr. 2, PLZ 91058 Erlangen, Germany}

$^3$ \textit{Institute for Theoretical Solid State Physics, RWTH Aachen University, 52074 Aachen, Germany}

$^4$ \textit{Max Planck Institute for the Science of Light, Staudtstrasse 2, 91058 Erlangen, Germany}

\end{center}

%%%%%%%%%% Merge with supplemental materials %%%%%%%%%%
%%%%%%%%%% Prefix a "S" to all equations, figures, tables and reset the counter %%%%%%%%%%
\setcounter{equation}{0}
\setcounter{figure}{0}
\setcounter{table}{0}
\setcounter{page}{1}
\makeatletter
\renewcommand{\theequation}{S\arabic{equation}}
\renewcommand{\thefigure}{S\arabic{figure}}
\renewcommand{\bibnumfmt}[1]{[S#1]}
\renewcommand{\citenumfont}[1]{S#1}
%%%%%%%%%% Prefix a "S" to all equations, figures, tables and reset the counter %%%%%%%%%%

\section{Experimental Setup}
The hybrid magnomechanical system used in this work is similar to those reported previously in Ref.~\cite{zhang2016cavity,potts2021dynamical,shen2022mechanical}. It consists of three coupled bosonic modes: a microwave electromagnetic field confined within a rectangular copper cavity, uniform magnetic excitations, and mechanical vibrations, the latter two hosted within a sphere of YIG. To begin, we can describe the system using these three interacting bosonic modes, as shown in Fig.~1(a), described by annihilation operators: $\hat{a}$ for the electromagnetic mode, $\hat{m}$ for the magnon mode, and $\hat{b}$ for the phonon mode, with frequencies $\omega_{\rm a}$, $\omega_{\rm m}$, and $\Omega_{\rm b}$, respectively. The Hamiltonian describing the hybrid magnomechanical system consists of two main coupling mechanisms: linear magnon-photon coupling \cite{huebl2013high,zhang2014strongly,tabuchi2015coherent} due to inductive coupling, and parametric magnon-phonon coupling \cite{keshtgar2014acoustic,callen1968magnetostriction} generated by magnetoelastic effects. In the frame rotating with the drive field, the Hamiltonian can be written in the form
\begin{equation}
\begin{aligned}
    \mathcal{H} =& {} -\hbar\Delta_{\textrm{a}} \hat{a}^{\dagger}\hat{a} - \hbar\Delta_{\textrm{m}} \hat{m}^{\dagger}\hat{m} + \hbar\Omega_{\rm b}\hat{b}^{\dagger}\hat{b} + \hbar g_{\text{am}}(\hat{a}\hat{m}^{\dagger} + \hat{a}^{\dagger}\hat{m}) + \hbar g_\text{mb}^0 \hat{m}^\dagger \hat{m}(\hat{b}^\dagger +\hat{b}) + i \hbar \sqrt{\kappa_{\rm{e}}} \epsilon_{\rm{d}} (\hat{a} -\hat{a}^\dagger).
\label{Hamiltonian01}
\end{aligned}
\end{equation}
The cavity and magnon detunings are defined as $\Delta_{\rm a} = \omega_{\rm d} - \omega_{\rm a}$ and $\Delta_{\rm m} = \omega_{\rm d} - \omega_{\rm m}$, respectively, where $\omega_{\rm d}$ is the drive frequency. The magnon-photon coupling rate is $g_{\rm am}$, the single-magnon magnomechanical coupling rate is $g_{\rm mb}^0$, the external coupling rate is $\kappa_{\rm e}$, and the external drive $\epsilon_{\rm d} = \sqrt{\mathcal{P}/\hbar\omega_{\rm d}}$, where $\mathcal{P}$ is the microwave drive power as measured at the microwave cavity, and $\hbar$ is the reduced Planck's constant.
\begin{figure}[b]
\includegraphics[width = 0.55\textwidth]{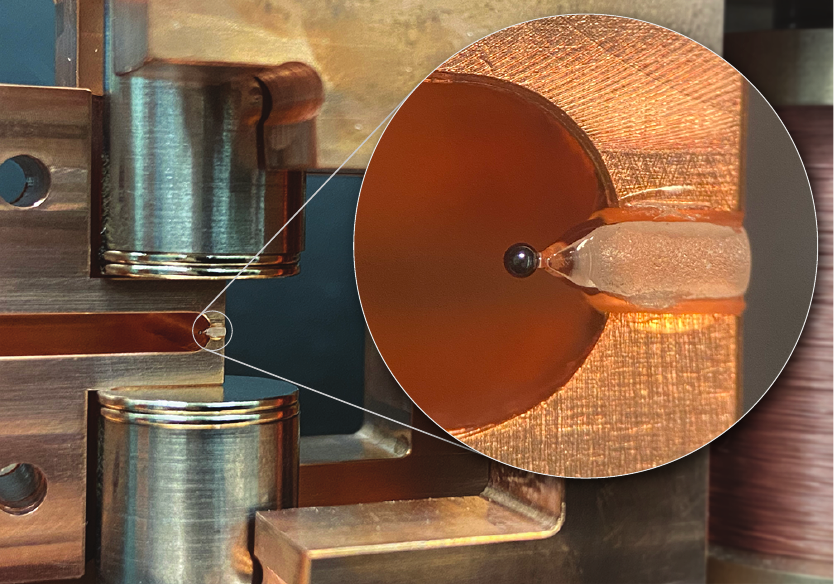}
\caption{Photograph of the experimental setup, similar to that used in Ref.~\cite{potts2021dynamical}. The YIG sphere was mounted on a sapphire stylus using UV-curing epoxy. A set of permanent neodymium magnets attached to a pure iron yoke provide the bias magnetic field, a solenoid allows the bias field to be varied dynamically. }
\label{Fig:App01}
\end{figure}

The linear magnon-microwave interaction results in the formation of normal (or hybrid) modes that are a superposition of magnons and photons. We label these modes as $+$ and $-$ for the upper and lower (frequency) normal modes, respectively. The frequency separation between the normal modes is given by
\begin{equation}
\label{eq:normalmodefreq}
    \Delta\omega \equiv \omega_+-\omega_- = \sqrt{4g_\text{am}^2 + \Delta_\text{am}^2},
\end{equation}
where $\Delta_{\rm am} = \omega_{\rm a} - \omega_{\rm m}$ is the difference between the magnon and photon frequencies, defined to be zero at maximal mode hybridization. The flexibility of this system can be seen by considering the fact that the magnon frequency can be controlled using an externally applied magnetic field. The frequency of the magnon mode is $\omega_{\rm m} = \gamma \vert \mathbf{B}_0 \vert$, where $\gamma/2\pi = 28$ GHz/T is the gyromagnetic ratio, and $\mathbf{B}_0$ is the externally applied static magnetic field. By varying the externally applied magnetic field, the magnon-photon detuning can be dynamically modified and thus, the normal mode splitting can be adjusted. The normal mode spectrum can be straightforwardly measured using a microwave signal. The inset of Fig.~2 shows the normal mode splitting for this experiment.

Our experimental setup consists of a three-dimensional microwave cavity machined from oxygen-free high-conductivity copper. The microwave resonator has inner dimensions $37 \times 26 \times 2.5\,$mm$^3$ and has a resonance frequency $\omega_{\rm a} = 7.11\,$GHz. The intrinsic decay rate of the microwave resonator is $\kappa_{\rm i}/2\pi = 2.91\,$MHz, and an external coupling rate $\kappa_{\rm e}/2\pi = 3.17\,$MHz, resulting in a total linewidth $\kappa / 2\pi = 6.08\,$MHz. A YIG sphere of 250 $\mu$m diameter was mounted using epoxy to a sapphire stylus and located at the magnetic field maximum of the microwave resonator. Due to the mounting, we observed that the phonon linewidth was larger than in previous experiments \cite{potts2021dynamical,zhang2016cavity}. The YIG sphere is positioned between a pair of neodymium magnets to generate a static bias magnetic field; the field can be tuned using a solenoid wrapped around a pure iron core, as described in Ref.~\cite{potts2021dynamical,tabuchi2015coherent}. The placement of the magnetic sphere did not take into account any particular relative orientation between the crystalline anisotropy axes of YIG and the other relevant directions: the external bias field and the magnetic field of the microwave cavity. The linewidth of the Kittel magnon mode is $\gamma_{\rm m}/2\pi = 2.55$ MHz. 

\section{Magnomechanical Damping Rate}

\begin{figure}[b]
\includegraphics[width = 0.95\textwidth]{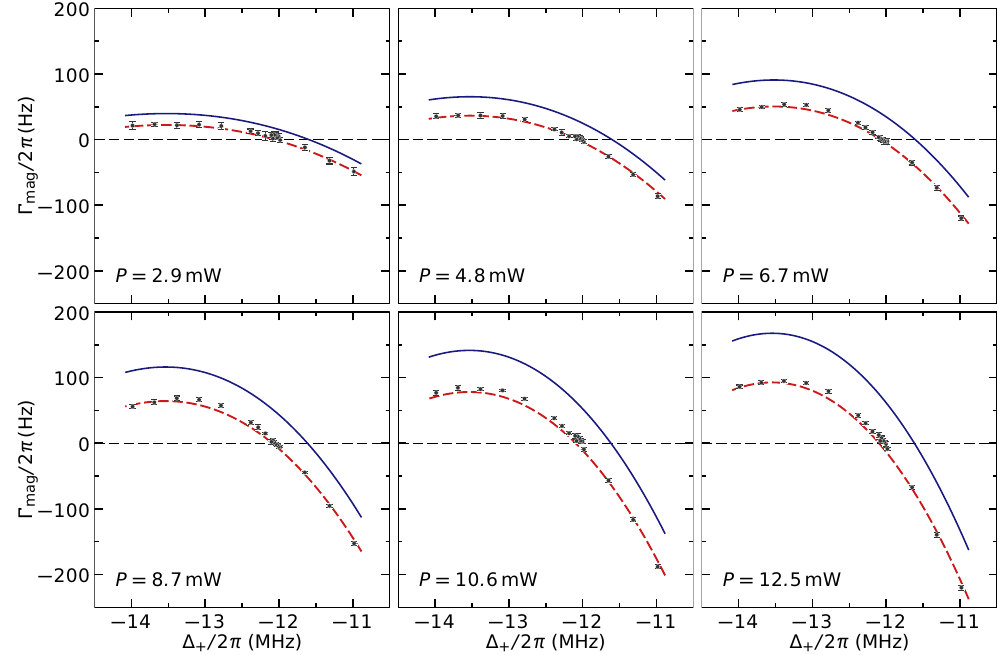}
\caption{Magnomechanical damping rate as a function of detuning for multiple drive powers within the linear regime. Fits were performed on all powers simultaneously, giving $g_{\rm mb}^0/2\pi = 4.56$ mHz and $\alpha/2\pi = -1.24$ pHz.}
\label{Fig:App01}
\end{figure}

As described in the main text, the magnomechanical damping rate was simultaneously fit for multiple drive powers. Here we show the magnomechanical damping rate as a function of drive detuning for multiple drive powers. In figure~\ref{Fig:App01} the blue curves represent the linear theory $\Gamma_{\rm mag} = 2 {\rm Im}\Sigma[\omega]$, and the red dashed curves are the modified linear theory given by,
\begin{equation}
    \Gamma_{\rm mag} = 2 {\rm Im}\Sigma[\omega] + \alpha \vert \langle \hat{m} \rangle \vert^2.
\end{equation}
Here $\Sigma[\omega]$ is the magnomechanical self-energy, $\vert \langle \hat{m} \rangle \vert^2$ is the steady-state magnon population, and $\alpha$ is a free fit parameter. We see a good agreement between the modified linear theory and the measured magnomechanical decay rate at all powers within the linear regime.

\section{Hybrid Magnon-Photon Modes}
\label{App:01}
Starting from the Hamiltonian given by Eq.~\eqref{Hamiltonian01}, we can define a transformation that diagonalizes the Hamiltonian \cite{zhang2016cavity}, given by,
\begin{equation}
    \begin{pmatrix}
    \hat{A}_+ \\
    \hat{A}_- 
    \end{pmatrix} = 
    \begin{pmatrix}
    \cos\theta & \sin\theta \\
    -\sin\theta & \cos\theta
    \end{pmatrix}
    \begin{pmatrix}
    \hat{a} \\
    \hat{m} 
    \end{pmatrix},
\end{equation}
where, as defined in the main text,
\begin{equation}
    \tan2\theta = \frac{-2g_{\rm am}}{\omega_{\rm a} - \omega_{\rm m}}.
\end{equation}
Written in terms of these newly defined modes, the Hamiltonian given by Eq.~\eqref{Hamiltonian01} may be written as,
\begin{equation}
\begin{aligned}
    \hat{\mathcal{H}} =&{}  -\hbar\Delta_+ \hat{A}_+^{\dagger}\hat{A}_+ - \hbar\Delta_- \hat{A}_-^{\dagger}\hat{A}_- + \hbar\Omega_b \hat{b}^{\dagger}\hat{b} \\ +&{} \hbar g^0_{\rm am}(\hat{b}+\hat{b}^{\dagger})[\sin^2\theta \hat{A}_+^{\dagger}\hat{A}_+ + \cos^2\theta \hat{A}_-^{\dagger}\hat{A}_- + \sin\theta\cos\theta (\hat{A}_+^{\dagger}\hat{A}_- + \hat{A}_-^{\dagger}\hat{A}_+)].
    \label{Ham:A1}
\end{aligned}
\end{equation}
Here $\hat{b}$ is the annihilation operator for the phonon mode, $\hat{A}_+$ and $\hat{A}_-$ are the annihilation operators for the upper and lower normal modes, respectively. From the Hamiltonian given by Eq.~\eqref{Ham:A1} we can derive the dynamics of a given operator $\hat{\mathcal{O}}$ using the Heisenberg equation $-i\hbar \dot{\hat{\mathcal{O}}} = [\hat{\mathcal{H}},\hat{\mathcal{O}}]$. Ignoring all quantum fluctuations, we find the semi-classical equations of motion for the expectation values $\langle \hat{A}_+ \rangle $, $\langle \hat{A}_- \rangle$ and $\langle \hat{b} \rangle$ are given by
\begin{equation}
\label{eq:App01Temp}
\begin{aligned}
 \langle \dot{\hat{A}}_+ \rangle &={}  (i\Delta_+ - \kappa_+/2)\langle \hat{A}_+ \rangle - ig_{mb}^0( \langle \hat{A}_+ \rangle \sin^2\theta +\langle \hat{A}_- \rangle \cos\theta\sin\theta)(\langle \hat{b} \rangle + \langle \hat{b}^{\dagger} \rangle) - \sqrt{\kappa_{\rm +,e}}\epsilon_{\rm d}, \\
 \langle \dot{\hat{A}}_- \rangle &={} (i\Delta_- - \kappa_-/2)\langle \hat{A}_- \rangle - ig_{mb}^0(\langle \hat{A}_- \rangle \cos^2\theta + \langle \hat{A}_+ \rangle \cos\theta\sin\theta)(\langle \hat{b} \rangle + \langle \hat{b}^{\dagger} \rangle) + \sqrt{\kappa_{\rm -,e}}\epsilon_{\rm d}, \\
 \langle \dot{\hat{b}} \rangle &={} (-i\Omega_{\rm b} - \Gamma_{\rm b}/2)\langle \hat{b} \rangle - ig_{mb}^0 \bigg[\vert  \langle \hat{A}_+\rangle\vert^2  \sin^2\theta  + \vert  \langle \hat{A}_-\rangle\vert^2\cos^2\theta  + \langle \hat{A}_+^{\dagger} \rangle \langle\hat{A}_- \rangle \sin\theta\cos\theta + \langle \hat{A}_-^{\dagger} \rangle \langle \hat{A}_+ \rangle\sin\theta\cos\theta \bigg]. 
\end{aligned}
\end{equation}
Here, $\kappa_+$ and $\kappa_-$ are the widths of the upper and lower normal modes, respectively, and we have ignored any quantum correlations such that, for example, $\langle \hat{A}_+^\dagger \hat{A}_- \rangle \approx \langle \hat{A}_+^\dagger \rangle \langle \hat{A}_- \rangle$. Next, to determine the classical steady-state population of the upper and lower normal modes, we set $\langle\dot{\hat{A}}_+\rangle = \langle\dot{\hat{A}}_-\rangle =0$. Solving for the steady-state amplitude and assuming $g_{mb}^0 \ll g_{\rm am}$, we find,
\begin{equation}
\begin{aligned}
    \langle \hat{A}_+ \rangle &= \frac{\sqrt{\kappa_{e,+}}\epsilon_{\rm d}}{i\Delta_+ - \kappa_+/2}, \\
    \langle \hat{A}_- \rangle &= -\frac{\sqrt{\kappa_{e,-}}\epsilon_{\rm d}}{i\Delta_- - \kappa_-/2}.
\end{aligned}
\end{equation}
Here, the external coupling rates for the upper and lower normal modes are given by
\begin{equation}
    \kappa_{ \rm{e},\pm} = \frac{1 \pm \cos2\theta}{2}\kappa_{\rm e}.
\end{equation}
The steady-state amplitude for the phonon mode $\langle \hat{b} \rangle$  is obtained by plugging the above solutions in \eqref{eq:App01Temp}.

We can now linearize the Hamiltonian by considering fluctuations about the steady-state values, letting $\hat{A}_{\pm} = \langle \hat{A}_{\pm} \rangle + \delta\hat{A}_{\pm}$ and  $\hat{b} = \langle \hat{b} \rangle + \delta\hat{b}$. Neglecting higher-order terms in fluctuations, we find the quadratic Hamiltonian
\begin{equation}
\begin{aligned}
    \hat{\mathcal{H}} &=  -\hbar\Delta_+ \delta\hat{A}_+^{\dagger}\delta\hat{A}_+ -\hbar \Delta_- \delta\hat{A}_-^{\dagger}\delta\hat{A}_- + \Omega_b \delta\hat{b}^{\dagger}\delta\hat{b} \\
    &\quad + \hbar g_{mb}^0 (\delta\hat{b}+ \delta\hat{b}^{\dagger})\bigg[ \langle \hat{A}_+ \rangle (\delta\hat{A}_+^{\dagger} + \delta\hat{A}_+)\sin^2\theta  + \langle \hat{A}_- \rangle(\delta\hat{A}_+^{\dagger} + \delta\hat{A}_+)\sin\theta\cos\theta \bigg] \\
    &\quad + \hbar g_{mb}^0 (\delta\hat{b} + \delta\hat{b}^{\dagger})\bigg[ \langle \hat{A}_- \rangle (\delta\hat{A}_-^{\dagger} + \delta\hat{A}_-)\cos^2\theta  + \langle \hat{A}_+ \rangle(\delta\hat{A}_-^{\dagger} + \delta\hat{A}_-)\sin\theta\cos\theta \bigg].
\end{aligned}
\end{equation}
This can be further simplified by defining two effective coupling rates,
\begin{equation}
    \begin{aligned}
        g_+ &= g_\text{mb}^0 [\langle \hat{A}_+ \rangle\sin^2\theta + \langle \hat{A}_- \rangle\sin (2 \theta) /2], \\
        g_- &= g_\text{mb}^0 [\langle \hat{A}_- \rangle\cos^2\theta + \langle \hat{A}_+ \rangle\sin (2 \theta) /2].
    \end{aligned}
\end{equation}
Using these definitions, we arrive at the Hamiltonian given in the main text
\begin{equation}
\begin{aligned}
    \hat{\mathcal{H}} =&{}  -\hbar\Delta_+ \delta\hat{A}_+^{\dagger}\delta\hat{A}_+ - \hbar\Delta_- \delta\hat{A}_-^{\dagger}\delta\hat{A}_- + \hbar\Omega_b \hat{b}^{\dagger}\hat{b} + \hbar g_+ (\delta\hat{A}_+^{\dagger} + \delta\hat{A}_+)(\delta\hat{b} + \delta\hat{b}^{\dagger})  + \hbar g_- (\delta\hat{A}_-^{\dagger} + \delta\hat{A}_-)(\delta\hat{b} + \delta\hat{b}^{\dagger}).
\end{aligned}
\end{equation}
For brevity, we dropped the $\delta$ in the Hamiltonian within the main text. Using the above Hamiltonian, we again derive the dynamics of an operator $\hat{\mathcal{O}}$  via the Heisenberg equation. For the normal modes operators and the phonon mode, the dynamics are given by the set of linear equations:
\begin{equation}
\begin{aligned}
    \dot{\delta\hat{A}}_+  &= (i\Delta_+ - \kappa_+/2)\delta\hat{A}_+ - ig_+(\delta\hat{b} + \delta\hat{b}^{\dagger}) + \hat{\xi}_+(t), \\
    \dot{\delta\hat{A}}_- &= (i\Delta_- - \kappa_+/2)\delta\hat{A}_- - ig_-(\delta\hat{b} + \delta\hat{b}^{\dagger})+ \hat{\xi}_-(t), \\
    \delta \dot{\hat{b}} &=  (-i\Omega_b - \Gamma_b/2)\hat{b} - i(g_+ \delta\hat{A}_+^{\dagger} + g_+^*\delta\hat{A}_+) \\&\quad -i(g_- \delta\hat{A}_-^{\dagger} + g_-^*\delta\hat{A}_-)+ \sqrt{\Gamma_{\rm{b}}}\hat{\xi}_{\rm{b}}(t).
    \label{Eqn:A10}
\end{aligned}
\end{equation}
In the above equations $\hat{\xi}_i(t)$ represents the noise terms. We can write Eq.~\eqref{Eqn:A10} in the frequency domain by performing a Fourier transform  $\delta\hat{\mathcal{O}}[\omega] = \int_{-\infty}^{\infty} dt e^{ i\omega t} \delta\hat{\mathcal{O}}(t)$:
\begin{equation}
\begin{aligned}
    \chi^{-1}_+[\omega] \delta\hat{A}_+[\omega] &= -ig_+(\delta\hat{b}[\omega] + \delta\hat{b}^{\dagger}[-\omega]) + \hat{\xi}_+[\omega], \\
    \chi^{-1}_-[\omega] \delta\hat{A}_-[\omega] &= -ig_-(\delta\hat{b}[\omega] + \delta\hat{b}^{\dagger}[-\omega]) + \hat{\xi}_-[\omega], \\
    \chi^{-1}_b[\omega] \delta \hat{b}[\omega] &= -i(g_+ \delta\hat{A}_+^{\dagger}[-\omega] + g_+^*\delta\hat{A}_+[\omega]) \\ &\quad  -i(g_- \delta\hat{A}_-^{\dagger}[-\omega] + g_-^*\delta\hat{A}_-[\omega])+ \sqrt{\Gamma_{\rm{b}}}\hat{\xi}_{\rm{b}}[\omega].
    \label{Eqn:A11}
\end{aligned}
\end{equation}
Here, $\chi^{-1}_{\pm}[\omega] = -i(\Delta_{\pm} + \omega) + \kappa_{\pm}/2$ is the effective susceptibility of the upper and lower normal modes, and $\chi^{-1}_{\rm{b}}[\omega] = [i(\Omega_{\rm{b}} - \omega) + \Gamma_{\rm{b}}/2]$. Solving this set of equations, we arrive at:
\begin{equation}
\begin{aligned}
    \delta\hat{b}[\omega][\chi_b^{-1}[\omega] &+\vert g_+ \vert^2 (\chi_+[\omega] - \chi_+^*[-\omega]) \\&-\vert g_- \vert^2 (\chi_-[\omega] - \chi_-^*[-\omega])] = \Upsilon.
\end{aligned}
\end{equation}
Where $\Upsilon$ represents all noise terms driving the phonon mode. From this equation, we can readily identify the self-energy as
\begin{equation}
\begin{aligned}
    \Sigma[\omega] &= i\vert g_+ \vert^2 (\chi_+[\omega] - \chi_+^*[-\omega])\\ &\quad + i\vert g_- \vert^2 (\chi_-[\omega] - \chi_-^*[-\omega]).
\end{aligned}
\end{equation}
The additional magnomechanical damping is defined as $2{\rm Im}\Sigma[\omega]$ \cite{aspelmeyer2014cavity,potts2021dynamical}. Ignoring counter-rotating terms, we can approximate the total magnomechanical damping rate as $2{\rm Im}\Sigma[\omega] \sim \Gamma_+ - \Gamma_-$, where,
\begin{equation}
    \begin{aligned}
    \Gamma_+ &\sim \frac{4\vert g_+ \vert^2 \kappa_+}{4(\Delta_+ + \Omega_b)^2+\kappa_+^2}, \\
    \Gamma_- &\sim \frac{4\vert g_- \vert^2 \kappa_-}{4(\Delta_- - \Omega_b)^2+\kappa_-^2}.
    \end{aligned}
\end{equation}
Therefore, we arrive at the total magnomechanical damping rate defined in the main text, given by
\begin{equation}
\label{eqn_3}
    \Gamma_{\rm mag} \simeq \Gamma_+ - \Gamma_-.
\end{equation}
We should note that this equation is an approximation, and the full expression is derived in Ref.~\cite{potts2020magnon}. However, based on numerical simulations, the approximate formula agrees well and provides an intuitive description of dynamical backaction evasion. 

Finally, the driving frequency which yields zero dynamical back action is obtained by solving $\Gamma_+ = \Gamma_-$, which is given by the following quadratic equation for $\omega_{\rm{d}}$
\begin{equation}
\begin{aligned}
&(\vert g_+ \vert^2 \kappa_+ -\vert g_- \vert^2 \kappa_-) \omega^2_{\rm{d}} - 2 [\vert g_+ \vert^2 \kappa_+(\omega_- + \Omega_{\rm{b}}) -\vert g_- \vert^2 \kappa_-(\omega_+ - \Omega_{\rm{b}})] \omega_{\rm{d}} \\
&\quad + \vert g_+ \vert ^2 \kappa_+ \left[(\omega_- + \Omega_{\rm{b}})^2 + \frac{\kappa_-^2}{4} \right] -\vert g_- \vert ^2 \kappa_- \left[(\omega_+ - \Omega_{\rm{b}})^2 + \frac{\kappa_+^2}{4} \right] =0.
\end{aligned}
\end{equation}
In the case of maximum hybridization, when the microwaves are resonant with the magnons (i.e. $\Delta_{\rm am}=0$), we have $g_+ = g_-$ and $\kappa_+ = \kappa_-$, and the above equation is linear in $\omega_{\rm{d}}$ with the solution
\begin{equation}
\omega_{\rm{d}} = \frac{\omega_+ + \omega_-}{2}.
\end{equation}
In other words, dynamical backaction is avoided by driving at a frequency exactly in the middle of the normal modes. If $\vert g_+ \vert^2 \kappa_+ \neq \vert g_- \vert^2 \kappa_-$, the driving frequencies for zero magnomechanical decay are
\begin{equation}
\begin{aligned}
\omega_{\rm{d}} &= \frac{\vert g_+ \vert^2 \kappa_+ (\omega_- + \Omega_{\rm{b}}) -\vert g_- \vert^2 \kappa_- (\omega_+ - \Omega_{\rm{b}})}{\vert g_+ \vert^2 \kappa_+  -\vert g_- \vert^2 \kappa_- } \\
&\quad \pm \frac{1}{\vert g_+ \vert^2 \kappa_+  -\vert g_- \vert^2 \kappa_-}\sqrt{\vert g_+ \vert^2 \vert g_- \vert^2 \kappa_+ \kappa_- \left[\left(\omega_+ - \omega_- - 2 \Omega_{\rm{b}} \right)^2 + \frac{\kappa_+^2+\kappa_-^2}{4} \right] - \frac{\kappa^2_+ \kappa^2_-}{4}\left( \vert g_+ \vert^4 + \vert g_- \vert^4 \right)}.
\end{aligned}
\end{equation}
In practice, this frequency is estimated numerically and found experimentally.

\section{\label{App:03}Single verses Two-phonon Triple Resonance}

\begin{figure*}[t]
\includegraphics[width = 0.90\textwidth]{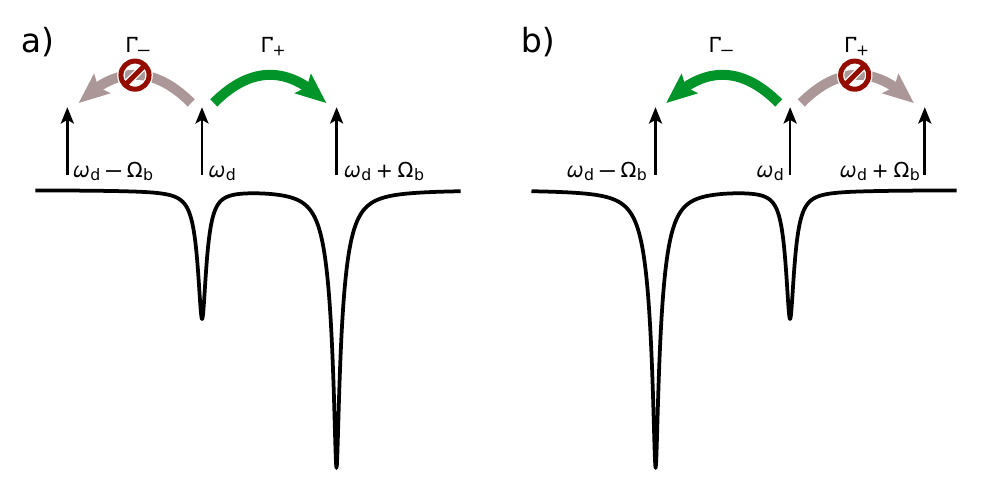}
\caption{Triple resonant cavity magnomechanics experiment. (a) When applying a drive tone resonant with the lower-normal mode, anti-Stokes scattering into the upper-normal mode is resonantly enhanced. Stokes scattering is far off-resonance and is therefore suppressed. (b) When applying a drive tone resonant with the upper-normal mode, Stokes scattering into the lower-normal mode is resonantly enhanced. Anti-Stokes scattering is far off-resonance and is therefore suppressed.}
\label{Fig:App03}
\end{figure*}

Finally, we wish to provide a short description outlining the difference between the two-phonon triple-resonance condition presented in this work and the tripe-resonance condition described in Ref.~\cite{potts2021dynamical}. The main text describes that the two-phonon triple resonance condition is met when the normal mode splitting equals twice the phonon frequency, that is, $2\Omega_{\rm b} = \Delta \omega$. In comparison, the triple-resonance condition is defined as when the normal mode splitting matches the phonon frequency, $\Omega_{\rm b} = \Delta \omega$. The triple-resonance condition allows for the selective enhancement of the Stokes or anti-Stokes scattering process depending on the drive detuning; see Fig.~\ref{Fig:App03} and Ref.~\cite{potts2021dynamical}. However, due to the finite cavity linewidth, evading dynamical backaction with a triply-resonant experimental setup is not possible. The different schemes are set by the frequency difference between the hybrid modes, which depends on the magnon-microwave coupling rate. The latter is, in turn, set by experimental design, so changing between the two aforementioned schemes would require a change in the experimental setup.

The inability to evade dynamical backaction can be understood by considering Fig.~\ref{Fig:App03}. When the microwave drive is resonant with the lower normal mode, the anti-Stokes scattering, $\Gamma_+$, is enhanced since excitations resonantly scatter into the upper normal mode. In contrast, since the experiment is sideband resolved, the Stokes process, $\Gamma_-$, is suppressed. A similar argument can be applied when the drive is tuned on resonance with the upper normal mode. Now the Stokes scattering process, $\Gamma_-$ is resonantly enhanced while the anti-Stokes process, $\Gamma_+$ is suppressed. In order to evade dynamical backaction, the condition described by Eq.~(3) must be met, that is, $\Gamma_+ = \Gamma_-$. Indeed this condition can be met by tuning the drive frequency between the two normal modes. However, in this detuning configuration, the red and blue sidebands are well outside the linewidth of the normal modes. Therefore, both scattering rates are suppressed such that it is not possible to observe the mechanical peak experimentally. The two-phonon triple resonance condition is required to improve the signal-to-noise ratio of the measurement by ensuring the mechanical sidebands scatter resonantly into the normal modes.  

\end{document}